\documentclass[a4paper,11pt]{article}
\usepackage{amsmath,amssymb,bm,mathrsfs,graphicx,rotating,feynmp,pstricks,mathtools,cancel}
\usepackage{caption,subcaption,multirow}
\usepackage{verbatim}
\usepackage{color}
\pdfoutput=1 

\usepackage{jheppub} 

\usepackage[T1]{fontenc} 
\usepackage[normalem]{ulem}

\newcommand{\beq}{\begin{equation}}
\newcommand{\eeq}{\end{equation}}

\renewcommand{\subsubsection}[1]{\addtocounter{subsubsection}{1}
\par\nobreak
\medskip
\nobreak
\noindent{\it \thesubsubsection.  #1 }
\par\nobreak\medskip\nobreak}

\def\lpar#1#2#3#4{\rlap{\raise#3\hbox{$\hskip#4#1\left\{\mbox{\phantom{\rule[0mm]{0mm}{#2}}}\right.$}}}
\def\rpar#1#2#3#4{\rlap{\raise#3\hbox{$\hskip#4\left\}#1\mbox{\phantom{\rule[0mm]{0mm}{#2}}}\right.$}}}

\title{\boldmath Phenomenology of a Long-Lived LSP with R-Parity Violation}
\author[a]{Csaba Cs\'aki,}
\emailAdd{csaki@cornell.edu}
\affiliation[a]{Department of Physics, LEPP, Cornell University, Ithaca, NY 14853, USA}

\author[a]{Eric Kuflik,}
\emailAdd{kuflik@cornell.edu}
\author[a]{Salvator Lombardo,}
\emailAdd{sdl88@cornell.edu}
\author[b]{Oren Slone,}
\emailAdd{shtangas@gmail.com}
\affiliation[b]{Raymond and Beverly Sackler School of Physics and Astronomy, Tel-Aviv University, Tel-Aviv
69978, Israel
}

\author[b]{Tomer Volansky}
\emailAdd{tomerv@post.tau.ac.il}

\abstract{We present the leading experimental constraints on supersymmetric models with R-parity violation (RPV) and  a long-lived lightest superpartner (LSP). We consider both the well-motivated dynamical RPV scenario as well as the conventional holomorphic RPV operators. Guided by naturalness, we study the cases of stop, gluino, and higgsino LSPs with several possible leading decay channels in each case. The CMS displaced dijet and the ATLAS multitrack displaced vertex searches have been fully recast, with all cuts and vertex reconstruction algorithms applied. Heavy charged stable particle searches by CMS are also applied. In addition, we consider representative bounds for prompt LSP decays that are directly applicable. Our main results are exclusion plots in the $m_{\rm LSP}-\tau_{\rm LSP}$ plane for the various scenarios. We find that the natural parameter space ($m_{\tilde{t}} <800$ GeV, $m_{\tilde{g}}<1500$ GeV, $m_{\tilde{H}}<800$ GeV) is excluded for a long-lived LSP ($\tau_{\rm LSP} \gtrsim 1$~mm). }

\begin{document}
\maketitle
\flushbottom

\section{Introduction\label{sec:Intro}}

For the past three decades, supersymmetry has been widely considered the top candidate for completing the Standard Model (SM) of particle physics. The expectation was that if supersymmetry provides  the solution to the naturalness problem, the superpartners should lie somewhat below the TeV-scale, the energy regime probed by the Large Hadron Collider (LHC). However,   Run-I of the LHC did not yield any evidence for supersymmetry~\cite{Aad:2014wea,Chatrchyan:2013iqa,Aad:2014qaa,Aad:2014mfk,Aad:2014lra}, challenging our preconception of naturalness.

LHC searches  for supersymmetry typically imply the superpartner masses in the 
minimal supersymmetric extension of the Standard Model (MSSM) to be above 1 TeV, rendering most of the parameter space unnatural. Thus if supersymmetry is to remain natural, it must manifest itself non-minimally. This has led to many new ideas of non-minimal models and renewed interest in older ideas, such as R-parity violation (RPV)~\cite{ Aulakh:1982yn,Hall:1983id,Ross:1984yg,Barger:1989rk,Dreiner:1997uz,Bhattacharyya:1997vv,Barbier:2004ez}. Many of the constraints from standard superpartner searches can be evaded if the assumption of exact  R-parity conservation is relaxed (see e.g.~\cite{Nikolidakis:2007fc,Csaki:2011ge, Barbier:2004ez, Krnjaic:2012aj, Franceschini:2013ne, Csaki:2013we, Krnjaic:2013eta, Brust:2012uf, Graham:2012th, FileviezPerez:2012mj, Han:2012cu, Berger:2012mm, Franceschini:2012za, Ruderman:2012jd} and references therein). In particular, if the LSP decays inside the detector, then supersymmetry searches which require large missing transverse energy (MET)  will not be applicable to these models.

At first sight the inclusion of RPV operators appears to lead to a new set of problems: each of the plethora of new couplings introduced has to remain small in order to suppress baryon- and lepton-violating processes. This suggests that there should be a systematic suppression of all RPV operators. A simple dynamical reason behind such a suppression would be if the visible sector (comprised of the SM fields and their superpartners) is R-parity conserving, while R-parity is violated in a hidden sector. The mediation of R-parity violation via heavy vector-like messengers will give rise to effective RPV operators in the visible sector, suppressed by the mass scale of the heavy messengers and possibly also by light fermion masses or the supersymmetry breaking scale.  Within this general framework, called dynamical R-parity violation (dRPV)~\cite{Csaki:2013jza,Csaki:2015fea}, all R-parity violation in the visible sector originates from effective higher dimensional operators. 

Since all RPV operators are dynamically suppressed, one needs to identify the leading RPV operators  in the visible sector. As shown in~\cite{Csaki:2013jza,Csaki:2015fea}, this identification depends on the details of the mediating dynamics and, in particular, on the  $U(1)_{B-L}$ and $U(1)_R$ charges of the fields responsible for the breaking of RPV.   One finds that while in some cases the standard holomorphic superpotential RPV  operators,
\beq
{\cal W}_{\rm RPV} =  \mu_i \ell_i h_u + \lambda_{ijk}  \ell_i\ell_j\bar{e}_k +\lambda^\prime_{ijk}  \ell_i q_j \bar{d}_k+\lambda^{\prime\prime}_{ijk} \bar{u}_i\bar{d}_j\bar{d}_k, \
\label{eq:WRPV}
\eeq
may dominate, typically the leading R-parity-violating operators will originate from the non-holomorphic, non-renormalizable K\"ahler terms,
\beq 
\label{eq:KRPV}
{\cal K}_{\rm nhRPV} = \frac{1}{M} \left( \eta_{ijk} \bar{u}_i\bar{e}_j\bar{d}_k^* + \eta^\prime_{ijk} q_i\bar{u}_j \ell_k^* +\eta^{\prime\prime}_{ijk} q_i q_j \bar{d}_k^* \right)\,. 
\eeq
The interactions from (\ref{eq:KRPV}) are suppressed either by the SUSY-breaking scale or by the fermion masses. The structure of the novel R-parity-violating operators (\ref{eq:KRPV}) and other non-standard RPV operators lead to new and distinct collider signatures. 

Both the standard RPV couplings (\ref{eq:WRPV}) and the non-holomorphic couplings (\ref{eq:KRPV}) must be small.  This often leads to LSP lifetimes of $c\,\tau_{\rm LSP} \gtrsim$~mm, corresponding to particles which are long-lived on collider scales. Thus the LSP can appear in the detector as an exotic non-prompt decay without necessarily producing missing energy. Both ATLAS and CMS have performed searches for such non-prompt decays, including searches for displaced vertices, decaying $R$-hadrons, and heavy stable charged particles (HSCP). 
The aim of this paper is to explore the final states allowed by the operators of~\eqref{eq:WRPV} and~\eqref{eq:KRPV}  and establish the resulting bounds in the LSP mass vs.~lifetime parameter space.  Our  focus is on  the cases motivated by naturalness, where the LSP is a stop squark, a gluino or a Higgsino.
We recast the main ATLAS~\cite{Aad:2015rba} and CMS~\cite{CMS:2014wda} searches for displaced vertices for the various R-parity violating scenarios. Some of the selected final state signatures are unique to the non-holomorphic operators (\ref{eq:KRPV}) (\textit{e.g.} $\tilde{t} \rightarrow \bar{b} \bar{b}$), while others have bounds which apply to both holomorphic and non-holomorphic RPV operators. HSCP searches by CMS are reinterpreted for the scenario where the LSP is unstable. We also present some results for   prompt searches by ATLAS and CMS to determine whether there exists unexplored gaps in the parameter space for short-lifetime LSPs.

While this work was nearing completion, a closely related analysis~\cite{Liu:2015bma}
by Zhen Liu and Brock Tweedie appeared which has a significant overlap with our paper.
 While our work studies the various RPV topologies, 
 \cite{Liu:2015bma} focuses on
  various supersymmetric scenarios. Our work also incorporates the recasting of the most recent ATLAS displaced vertex search~\cite{Aad:2015rba} and  uses slightly different tracking efficiencies and vertex reconstruction procedures.

The   paper is structured as follows.  In Sec.~\ref{sec:Theory} we present the models and parameter space which we consider for our analysis.  In Sec.~\ref{sec:Searches} we briefly describe the experimental searches considered in our analysis.  In Sec.~\ref{sec:Results} we present our results in the form of exclusion plots in the LSP mass vs.~lifetime parameter space. We conclude in Sec.~\ref{sec:Conclusions}. In Apps.~\ref{sec:Data} and \ref{sec:DetailedSearches} we detail the analysis techniques implemented in the various experimental searches we have recast and present our procedure used for each channel.

\section{Theory and Models \label{sec:Theory}}

The aim of this project is to explore the RPV scenarios that are most relevant for maintaining a natural Higgs sector within the supersymmetric theory. This requires the stop be well below the TeV scale, the gluino lie not too far above the stop, and the Higgsinos to be light as well~\cite{Papucci:2011wy}. Thus we study models with light stops, gluinos, or Higgsinos below the TeV scale.  The LHC phenomenology will strongly depend on the identity of the LSP and the nature of the RPV operator responsible for its decay.  In accordance with the naturalness expectations, we consider three types of LSP particles:  stops ($\tilde t$), gluinos ($\tilde g$), and Higginos ($\tilde H$).   

Since a
  main focus of this paper is the study of the non-holomorphic RPV operators of the dRPV scenario in (\ref{eq:KRPV}), we concentrate on decays that predict final outgoing jets (in addition to other possible final-state particles).  The reason for this is that the SM gauge structure of the non-holomorphic operators (\ref{eq:KRPV}) differs from the holomorphic operators (\ref{eq:WRPV}). In particular there is no non-holomorphic cubic term that contains more than one lepton (there is no analog of the $ll\bar{e}$ coupling), thus a jet is expected in every final state involving these operators. An exception which we do not consider could be the case  for a Higgsino LSP  if the breaking of R-parity occurs via bi-linear terms or via the holomorphic $ l l \bar e$ operator.

\begin{table}[t!]
 \begin{center}
\begin{tabular}{| c | c | c |  c |}
\hline
\multicolumn{4}{ |c| }{Topologies} \\
\hline
LSP &  Decay & Operator & Results \\ \hline
\multirow{3}{*}{$\tilde{t}$}  
 & $\bar{d}\, \bar{d}^\prime$ & $\lambda^{\prime\prime}$, $\eta^{\prime\prime}$ & Fig.~\ref{stopdd}  \\
 & $u\, \bar{\nu}$ & $\eta^\prime$ & Fig.~\ref{stopdl}  \\
 & $d\, \ell^+ $ & $\lambda^\prime$, $\eta$ & Fig.~\ref{stopun}  \\ \hline
\multirow{3}{*}{$\tilde{g}$} 
 & $t \, d \, d^\prime+c.c$  &$\lambda^{\prime\prime}$, $\eta^{\prime\prime}$&  Fig.~\ref{gluino1}  \\
 & $t\,\bar{u}\, \bar{\nu}+c.c$ & $\eta^\prime$ & Fig.~\ref{gluino1}  \\
 & $t \bar{d}\, \ell^- +c.c$ & $\lambda^\prime$, $\eta$&  Fig.~\ref{gluino2}  \\ \hline 
\multirow{3}{*}{$\tilde{H}^0/\tilde{H}^\mp$} 
& $(t/b) \, d \, d^\prime+c.c$  & $\lambda^{\prime\prime}$, $\eta^{\prime\prime}$&  Fig.~\ref{higgsino} \\
 & $(t/b)\,\bar{u}\, \bar{\nu}+c.c$ & $\eta^\prime$ &  Fig.~\ref{higgsino} \\
 & $(t/b)\, \bar{d}\, \ell^- +c.c$ &$\lambda^\prime$, $\eta$&   Fig.~\ref{higgsino} \\ \hline
\end{tabular}
 \end{center}
\caption{Summary of the various LSPs and  their 
decay channels considered in this paper. The third column denotes the RPV operators from (\ref{eq:WRPV}) or (\ref{eq:KRPV}) that can give rise to the decays and the final column points to the relevant exclusion plot in Section~\ref{sec:Results}.\label{Tab:summarychannels}}
\end{table}

For the various LSPs, the decay channels considered in this paper are summarized in Table~\ref{Tab:summarychannels}.  We only consider direct production of LSPs.
We study three possible decay channels for the stop:
{\begin{itemize}
\item ${ \tilde t\to \bar d \bar d^\prime}$  where $\bar d$, $\bar d^\prime$ are down-type SM quarks.   Such decays arise either from the $\lambda^{\prime\prime}$  holomorphic RPV operator in Eq.~\eqref{eq:WRPV} or from the $\eta^{\prime\prime}$ operator of the non-holomorphic dRPV operators in Eq.~\eqref{eq:KRPV}.  While  decays to only third generation particles $\tilde t \to \bar b \bar b$ are not allowed with   the holomorphic $\lambda''$ operator due to the antisymmetry of the color contraction, such decays are allowed and are expected to dominate in the case of the $\eta''$ operator in Eq.~\eqref{eq:KRPV} .   
\item $\tilde t \to u\bar\nu$ where $u$ is an up-type quark and $\bar{\nu}$ is an anti-neutrino.  This channel occurs for the $\eta^\prime$ operator of Eq.~\eqref{eq:KRPV} but is suppressed by the neutrino mass.  However, this decay can be induced by effective Kahler potential terms of the form $\mathcal{K} \supset  \left(\mathcal{D}^2 q \bar{u} \right) \ell^*$, which were shown in \cite{Csaki:2015fea} to be generated in a simple model of dRPV.  The decay can also be induced via superpotential terms of the form $\mathcal{W} \supset \ell h_u h_u q \bar{u}$ and via neutrino-higgsino mixing.
\item $\tilde t \to d l^+$ where $l^+$ is an anti-lepton.   The responsible operator may be $\lambda^\prime$ of Eq.~\eqref{eq:WRPV} or $\eta$ of Eq.~\eqref{eq:KRPV}.   
\end{itemize}}

The chiral suppression of the decays from the non-holomorphic operators and the typical flavor structure of RPV operators strongly suggests that stop decays to heavy quarks dominates over decays to light quarks.  Thus we expect that third generation final states will dominate both for the case of stop LSP and for the other possibilities discussed below.
Since the stop is assumed to be the lightest squark, the gluino decays via an off-shell stop and the final states will correspond to the above listed final states for stop decay with an additional top quark, $\tilde g \to t d d^\prime$ + c.c., $\tilde g \to t \bar u \bar \nu$ + c.c. and $\tilde g \to t \bar d l^-$ + c.c., with third generation final states again expected to dominate. 
Similarly, a neutral Higgsino LSP  
decays via an off-shell stop and its topologies are identical to those of the gluino LSP decays.  A charged Higgsino will instead have a bottom quark replacing the top quark in the respective final states. 

\section{Experimental Searches\label{sec:Searches}}

In this section, we give a brief overview of the searches we have used for our study. A more detailed discussion of the individual searches, as well as our methods of applying them to RPV models, can be found in Appendices~\ref{sec:Data} and~\ref{sec:DetailedSearches}. 
 
The LSP may decay  promptly at the primary vertex (PV) or may traverse a finite distance before decaying, depending on the size of the corresponding RPV coupling. While we mainly focus on displaced LSP decays due to RPV interactions, we consider the limits obtained from prompt and HSCP studies as well (corresponding to very short or long LSP lifetimes respectively).  

\subsection{Searches for Displaced Vertices}
LSPs which decay in the inner detector can be reconstructed as displaced vertices by both the ATLAS and CMS trackers. The SM background for these searches is
negligible. Long lifetimes, for which only a few of the produced LSPs decay within the tracker, can be excluded, ranging from 1 to 10$^7$ mm proper lifetimes. 
The searches presented in this paper look for displaced vertices associated with one of several signatures used for triggering, including jets, leptons, and MET. 
We recast displaced searches by both ATLAS~\cite{Aad:2015rba} and CMS~\cite{CMS:2014wda}, which together place strong constraints on all of the final states considered in this paper. The full detail of the simulation and procedure, including cuts, tracking efficiencies, and vertex reconstruction, are given in Appendices~\ref{sec:Data} and~\ref{sec:DetailedSearches}. \\

\noindent \textbf{ATLAS DV+$\mu / e /$jets/MET}.

\noindent The ATLAS search for long lived particles at $\sqrt{8} \text{ TeV}$ \cite{Aad:2015rba} considers four multitrack signatures with a luminosity of $20.3 \ \text{fb}^{-1}$ in which a displaced vertex (DV) with at least 5 tracks is accompanied by one of the following: (1) a high $p_T$ muon (2) a high $p_T$ electron (3) jets or (4) missing transverse energy (MET). These searches are background free. The muon search has been improved from the previous study at 8 TeV \cite{TheATLAScollaboration:2013yia} by increasing the detector volume used for reconstructing vertices.

\vspace*{0.3cm}
\noindent \textbf{CMS Displaced Dijet}.

\noindent The CMS collaboration has performed a search for dijets originating from displaced vertices at $\sqrt{s} = 8$ TeV with 18.6 fb$^{-1}$ of data \cite{CMS:2014wda}. The analysis requires two hard jets consistent with a dijet produced at a secondary displaced vertex. The constraints from this search are not limited to models with dijets. Isolated leptons are treated as jets, and three-jet final states have significant efficiency to be reconstructed as dijets, giving sensitivity to a variety of models.

\subsection{Searches for Stable Particles}
Long-lived colored particles, such as squarks or gluinos, form composite objects with SM particles, called $R$-hadrons.
For long enough lifetimes, the LSP may hadronize and traverse the whole detector before decaying.  Searches for  slowly moving charged particles constrain the production rate of a stable LSP. These searches also place strong limits on unstable LSPs with lifetimes long enough for a fraction of the particles to traverse the detector before decaying.  
\\

\newpage

\noindent \textbf{CMS Heavy Stable Charged Particle Search}.

\noindent CMS has performed a search for heavy stable charged particles (HSCP) at $\sqrt{s} = 8$ TeV with an integrated luminosity of 18.8 fb$^{-1}$ \cite{Chatrchyan:2013oca}. HSCPs have velocities much less than the speed of light leading to misidentification in the particle flow reconstruction algorithm used by CMS which assumes SM particles with $v\approx c$.  Additionally, tracks associated to the HSCP typically have large energy loss ($dE / dx$) due to interactions with the tracker. Thus, HSCPs are identified by a longer time-of-flight to the muon system or by tracks with anomalous energy loss.  $R$-hadrons can be either charged or neutral, but interactions with the detector material may   "charge-strip" the charged $R$-hadrons, converting them into neutral hadrons by the time they reach the outer muon system. To account for this possibility, CMS considers two models of the R-hadrons' strong interactions with the material in the detector. The first, referred to as the cloud-model, assumes that the HSCP is surrounded by a cloud of light SM colored particles. The second is a charge-suppressed model, in which charged $R$-hadrons quickly become neutral by interacting with the detector material and are often neutral before reaching the muon system. 

\subsection{Prompt Searches}

These searches are relevant for LSPs with small proper lifetimes, taken conservatively to be below 1 mm in this study, for which the decay products of the LSP point back to the primary production vertex. We have not recast
 prompt searches. Instead, we have used
existing bounds which are directly applicable to the particular scenarios under consideration. The searches used include the paired dijet resonance search at CMS~\cite{Khachatryan:2014lpa} for the fully hadronic prompt stop decays, leptoquark searches~\cite{Acosta:2005ge,Abazov:2009ab,CMS:2014qpa,Abulencia:2005ua,Abazov:2008np,CMS:zva,Aaltonen:2007rb,Abazov:2010wq,Khachatryan:2014ura} from the Tevatron and the LHC for $\tilde{t}\to dl^+$ decays, R-parity conserving supersymmetric searches~\cite{Aad:2014wea,Chatrchyan:2013iqa,Aad:2014qaa,Aad:2014mfk,Aad:2014lra,Aad:2015gna} for $\tilde{t}\to t\tilde{N}$, $\tilde{c}\to c\tilde{N}$, $\tilde{g}\to t\bar{t} \tilde{N}$ (for vanishing neutralino mass) for decays involving neutrinos, and the R-parity violating ATLAS gluino search~\cite{Aad:2014pda}.

\section{Results\label{sec:Results}}

In this section we present our results for the bounds on the LSP masses and lifetimes in the various scenarios considered and summarised in Table~\ref{Tab:summarychannels}. 
Our results follow from a detailed MC study which includes all channels and the application of
the cuts and the reconstruction procedures, designed to correspond to the various experimental displaced vertex searches.  We postpone the details and discussion of these procedures to Appendix~\ref{sec:Data}. 
 The HSCP search was recast at the parton level in order to find the efficiency for an unstable LSP to traverse the detector before decaying. Supersymmetric production cross sections were obtained from the LHC SUSY cross section working group \cite{lhcsusy,Kramer:2012bx} when applicable. Higgsino direct production was calculated at tree-level in Madgraph, and an NLO k-factor of 1.6 was applied for all masses \cite{Fuks:2012qx,Fuks:2013vua}.  Exclusion bounds from prompt searches are directly taken from the experimental results, and were not recast, implying that only a small fraction of the searches relevant to prompt decays have been included. To obtain further bounds in the prompt case requires recasting those searches and is beyond the scope of this paper.

\subsection{Stop LSP}
\begin{figure}[t!]
\center
\includegraphics[width=.48\textwidth]{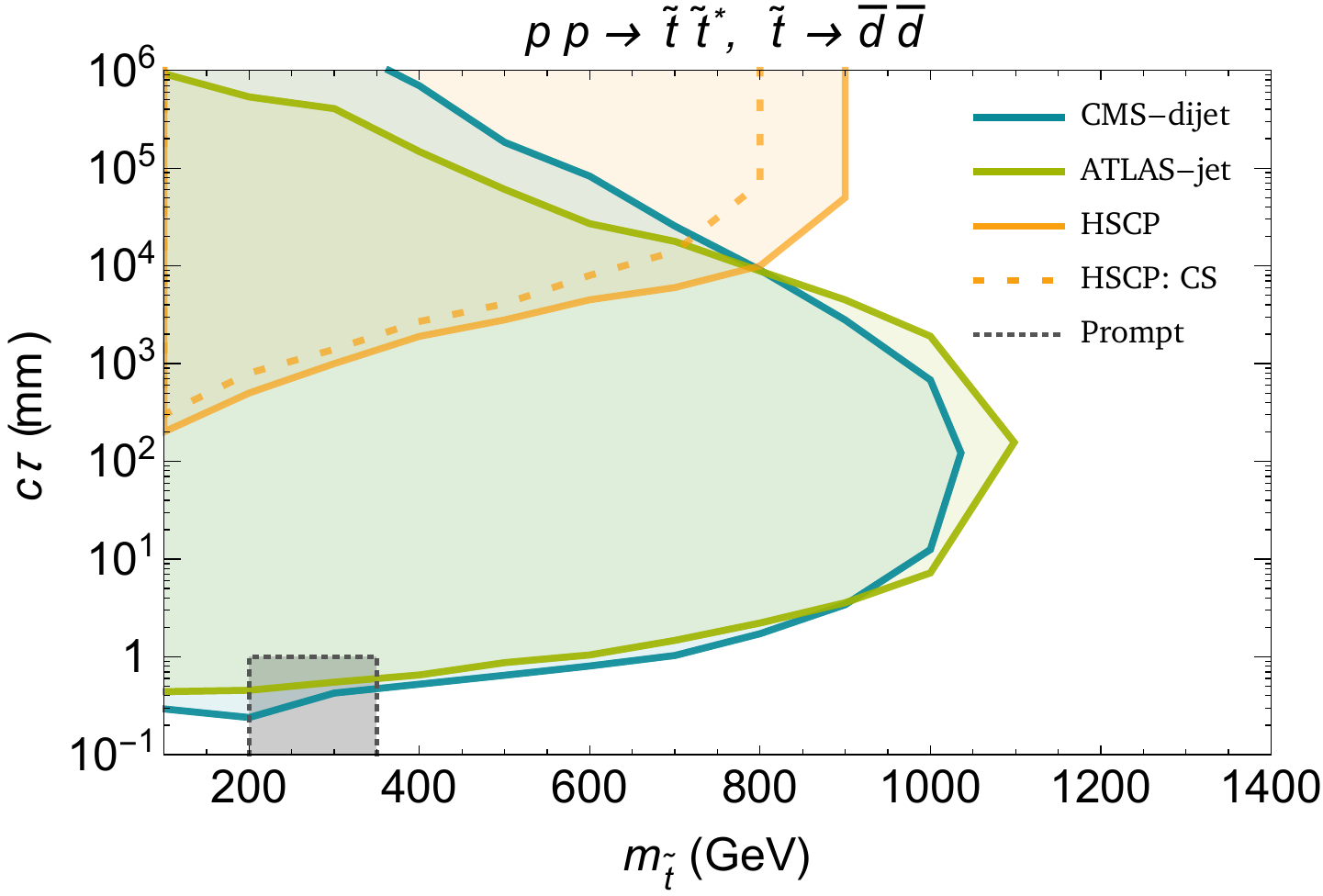}
\includegraphics[width=.48\textwidth]{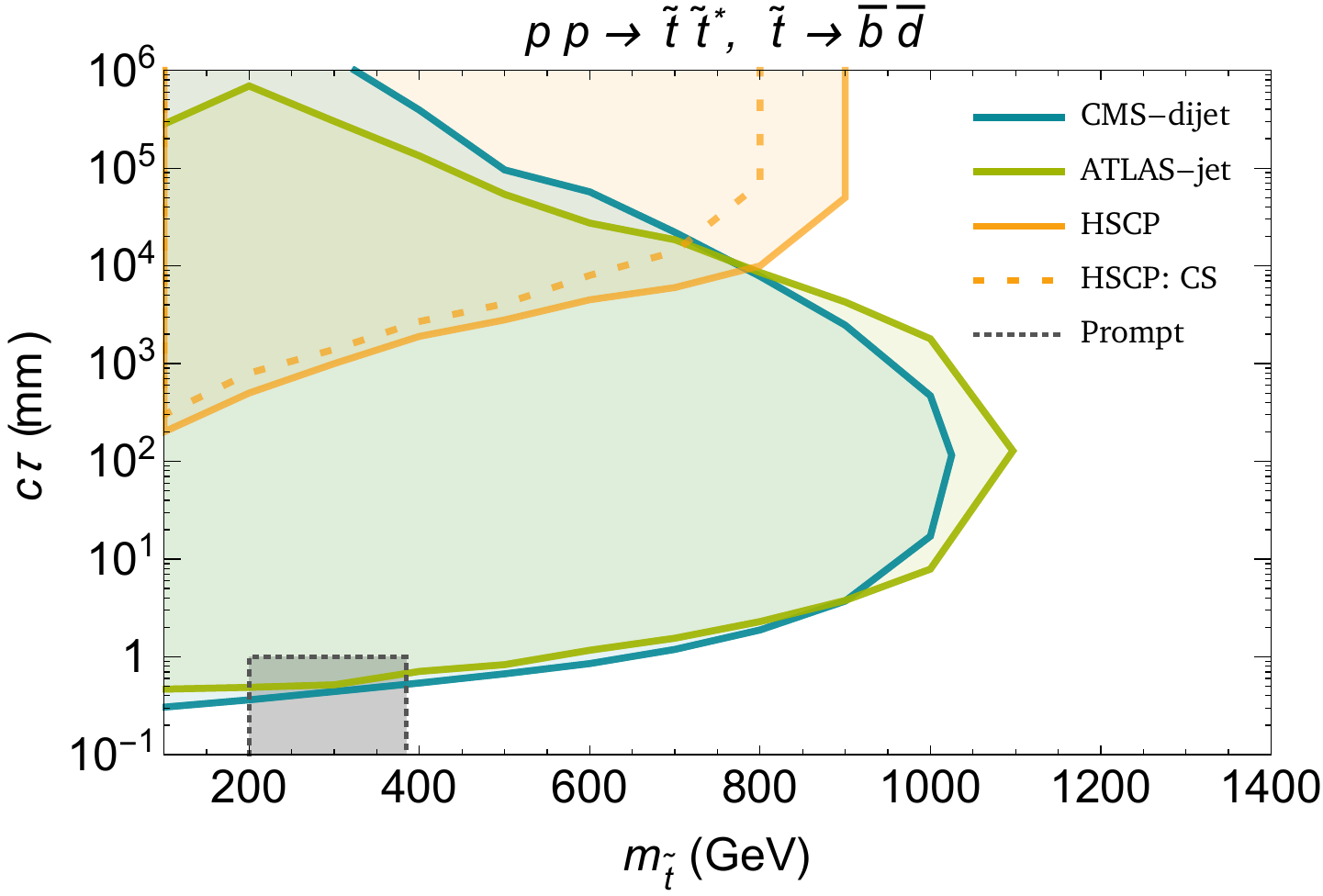}
\includegraphics[width=.48\textwidth]{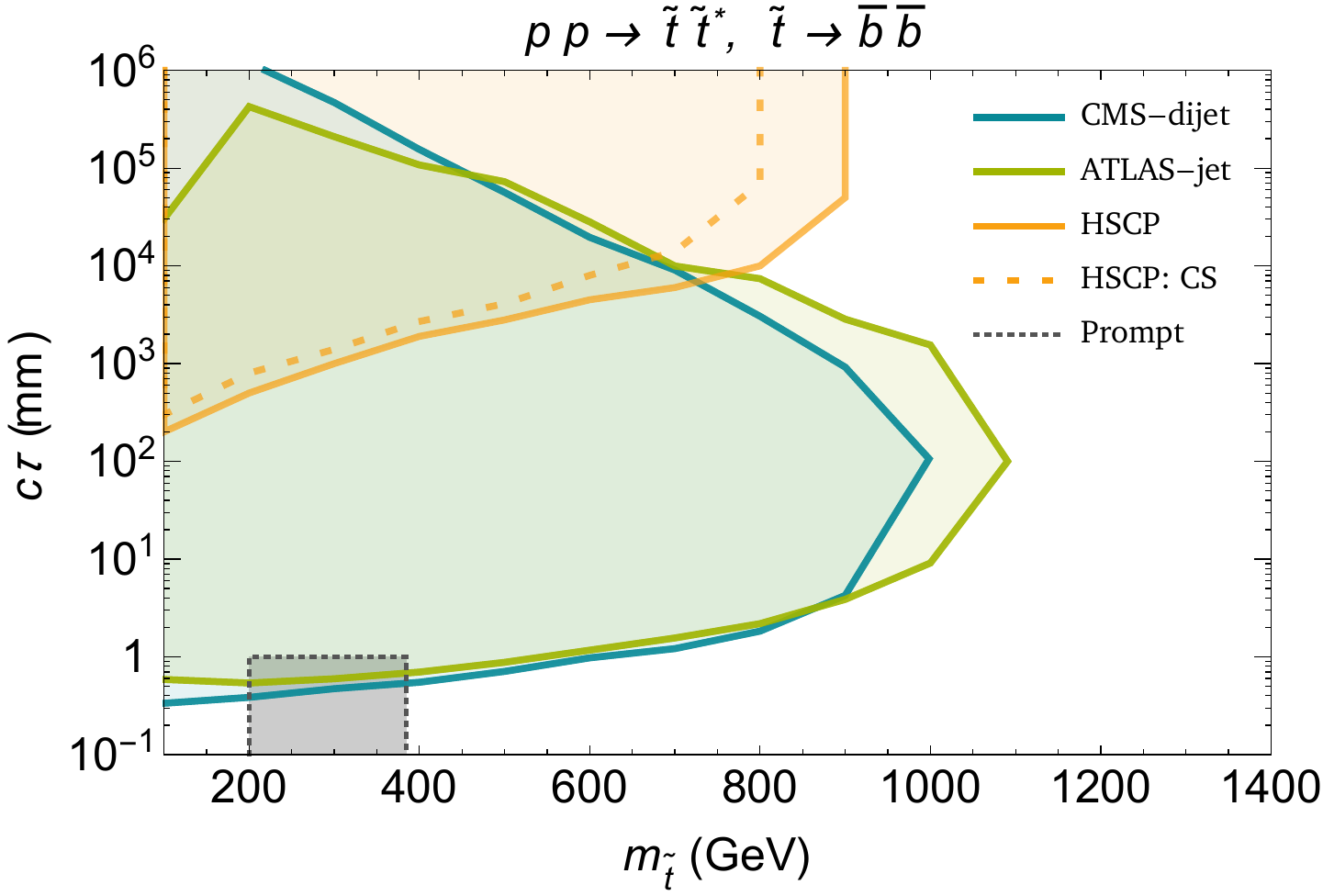}
\caption{95\% CL exclusion curves for fully hadronic stop decays to two down-type quarks. In the {\bf upper left} plot we show $\tilde{t}\to \bar{d}\bar{d}$, in the {\bf upper right} plot $\tilde{t}\to \bar{d}\bar{b}$, and in the {\bf bottom plot} $\tilde{t}\to \bar{b}\bar{b}$ is presented.  While all displaced searches of ATLAS~\cite{Aad:2015rba} and CMS~\cite{CMS:2014wda} have been recast, we only display the curves yielding the strongest constraints. Stable searches for HSCPs \cite{Chatrchyan:2013oca} were recast for an unstable R-hadron. The prompt searches here correspond to the CMS dijet resonance search~\cite{Khachatryan:2014lpa}.  \label{stopdd}}
\end{figure}

\begin{figure}[t!]
\center
\includegraphics[width=.48\textwidth]{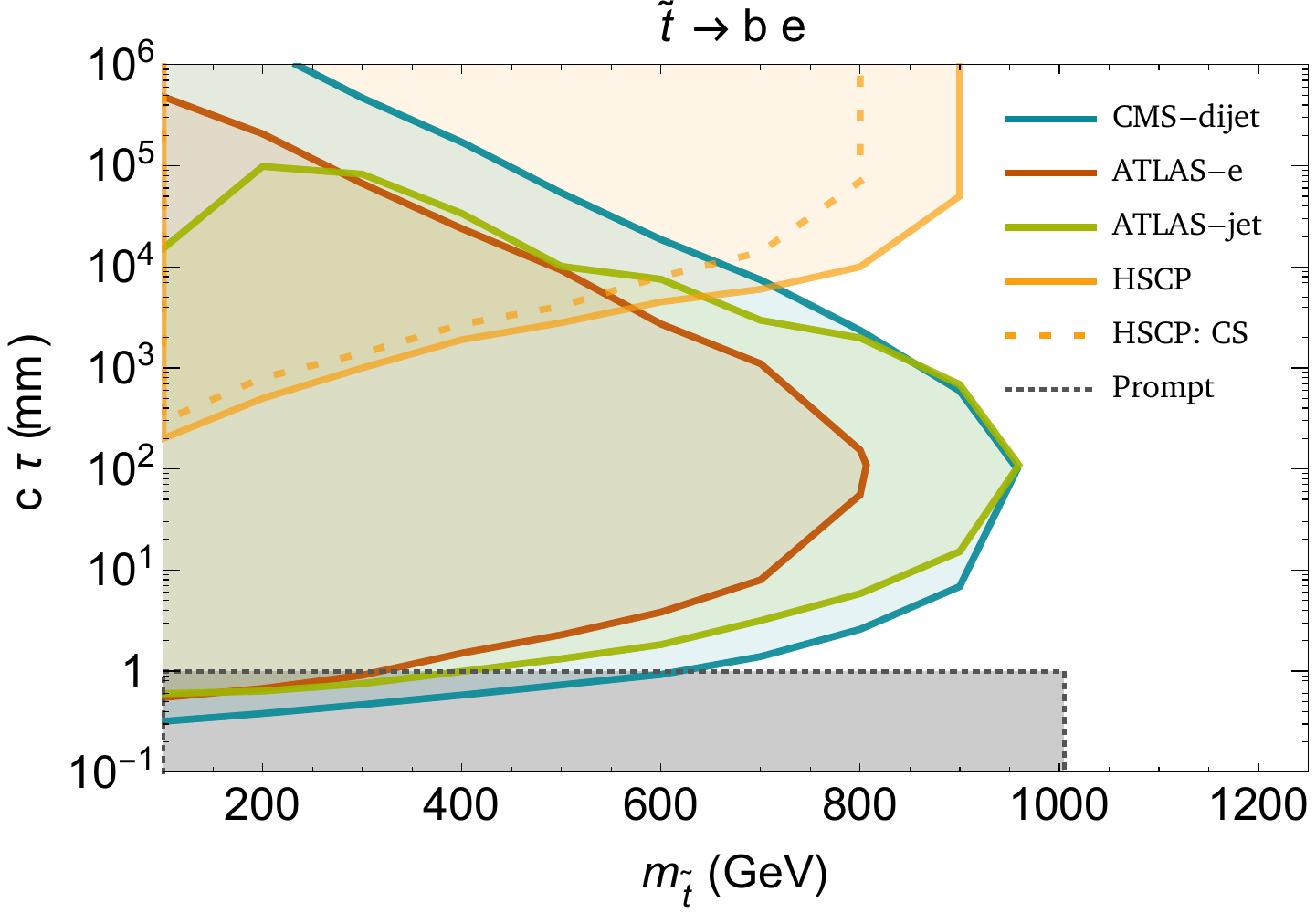}
\includegraphics[width=.48\textwidth]{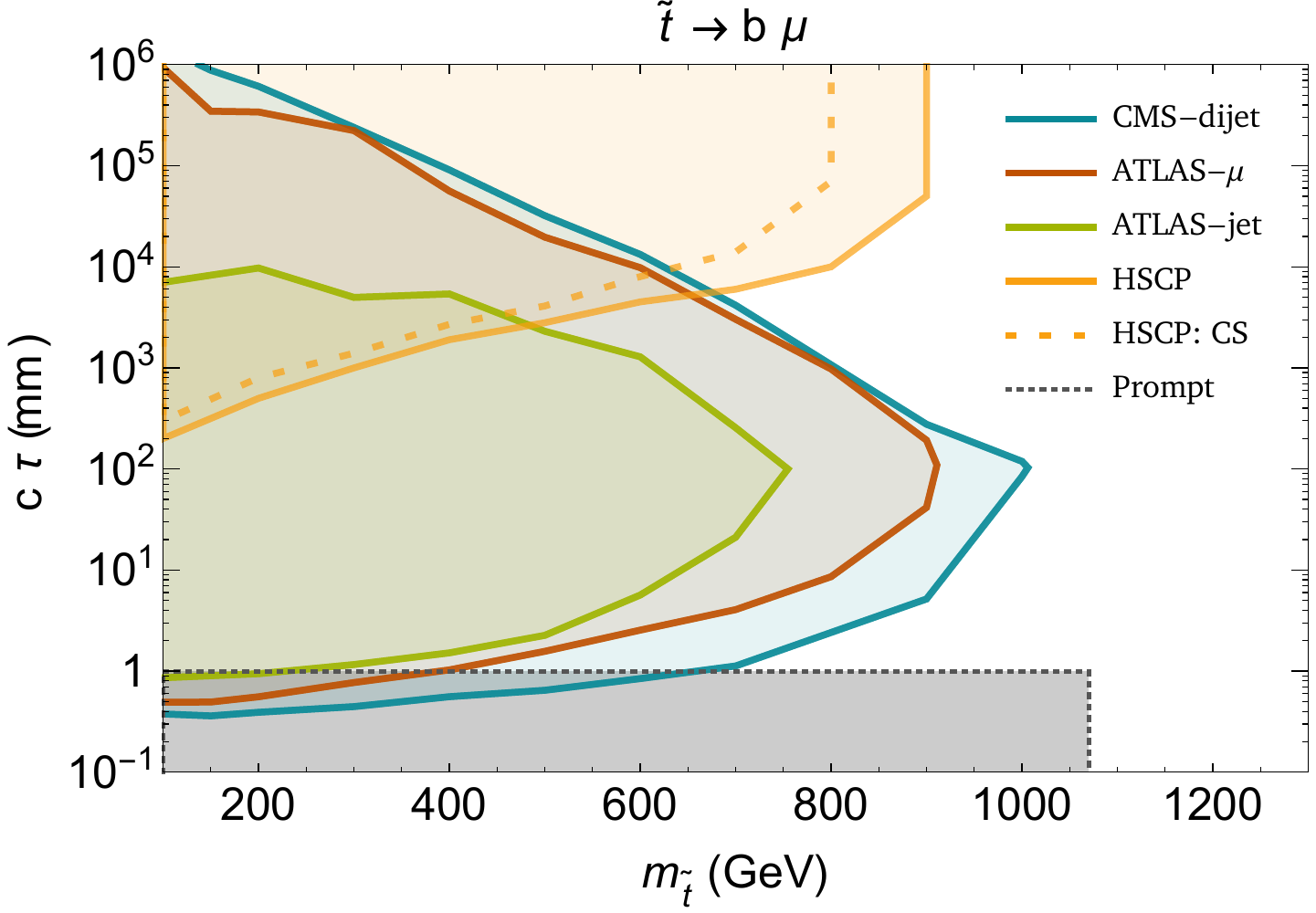}
\includegraphics[width=.48\textwidth]{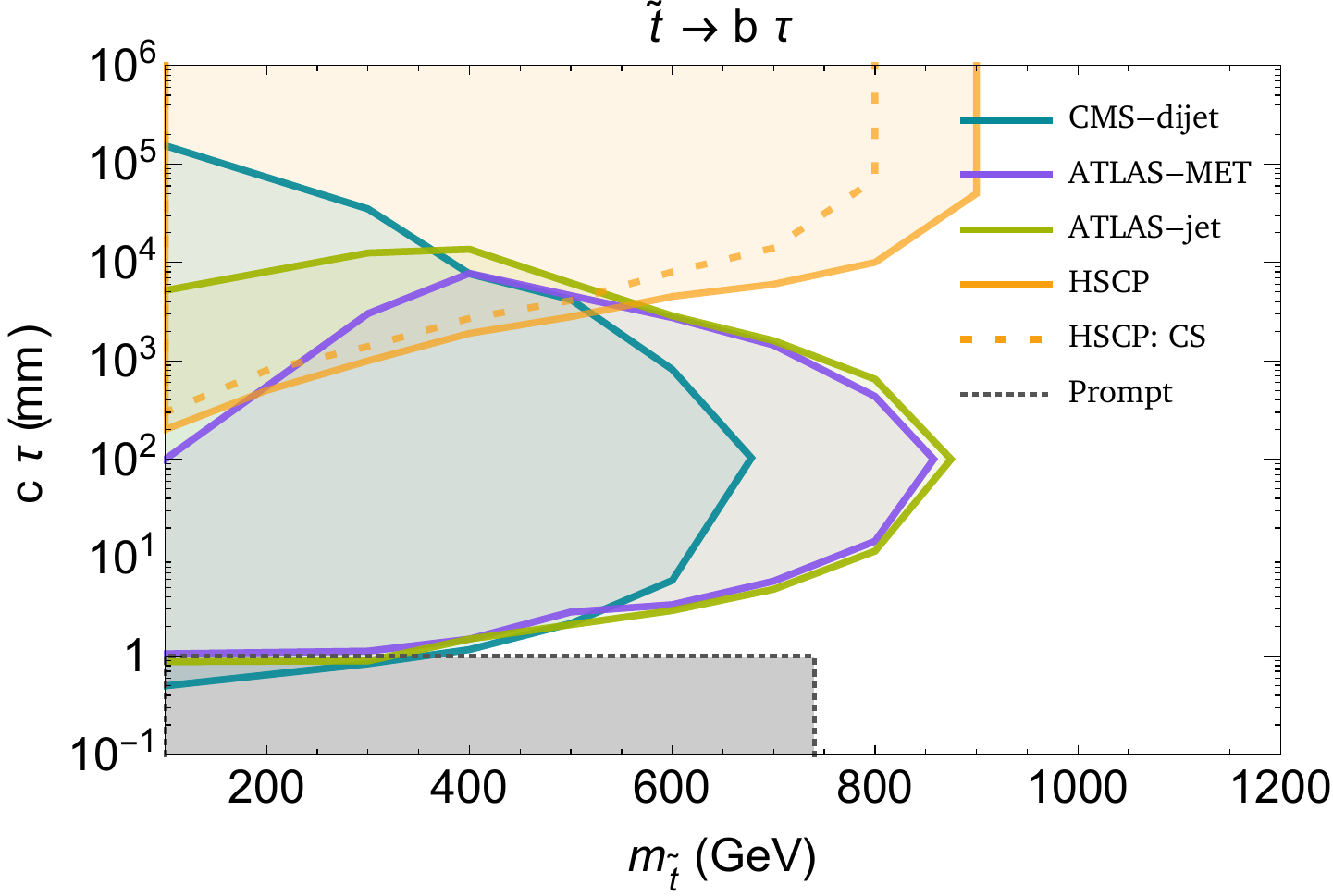}
\caption{95\% CL exclusion curves for stop decays to a bottom quark and a charged lepton. In the {\bf upper left}  plot we show $\tilde{t}\to be^+$, in the {\bf upper right}  plot $\tilde{t}\to b\mu^+$, and in the {\bf bottom plot} we have $\tilde{t}\to b\tau^+$. While all displaced searches of ATLAS~\cite{Aad:2015rba} and CMS~\cite{CMS:2014wda} have been recast, we only display the curves yielding the strongest constraints. Stable searches for HSCPs \cite{Chatrchyan:2013oca} were recast for an unstable R-hadron.  The prompt searches here correspond to leptoquark searches at the Tevatron and LHC~\cite{Acosta:2005ge,Abazov:2009ab,CMS:2014qpa,Abulencia:2005ua,Abazov:2008np,CMS:zva,Aaltonen:2007rb,Abazov:2010wq,Khachatryan:2014ura}. \label{stopdl}}
\end{figure}

\begin{figure}[ht!]
\center
\includegraphics[width=.48\textwidth]{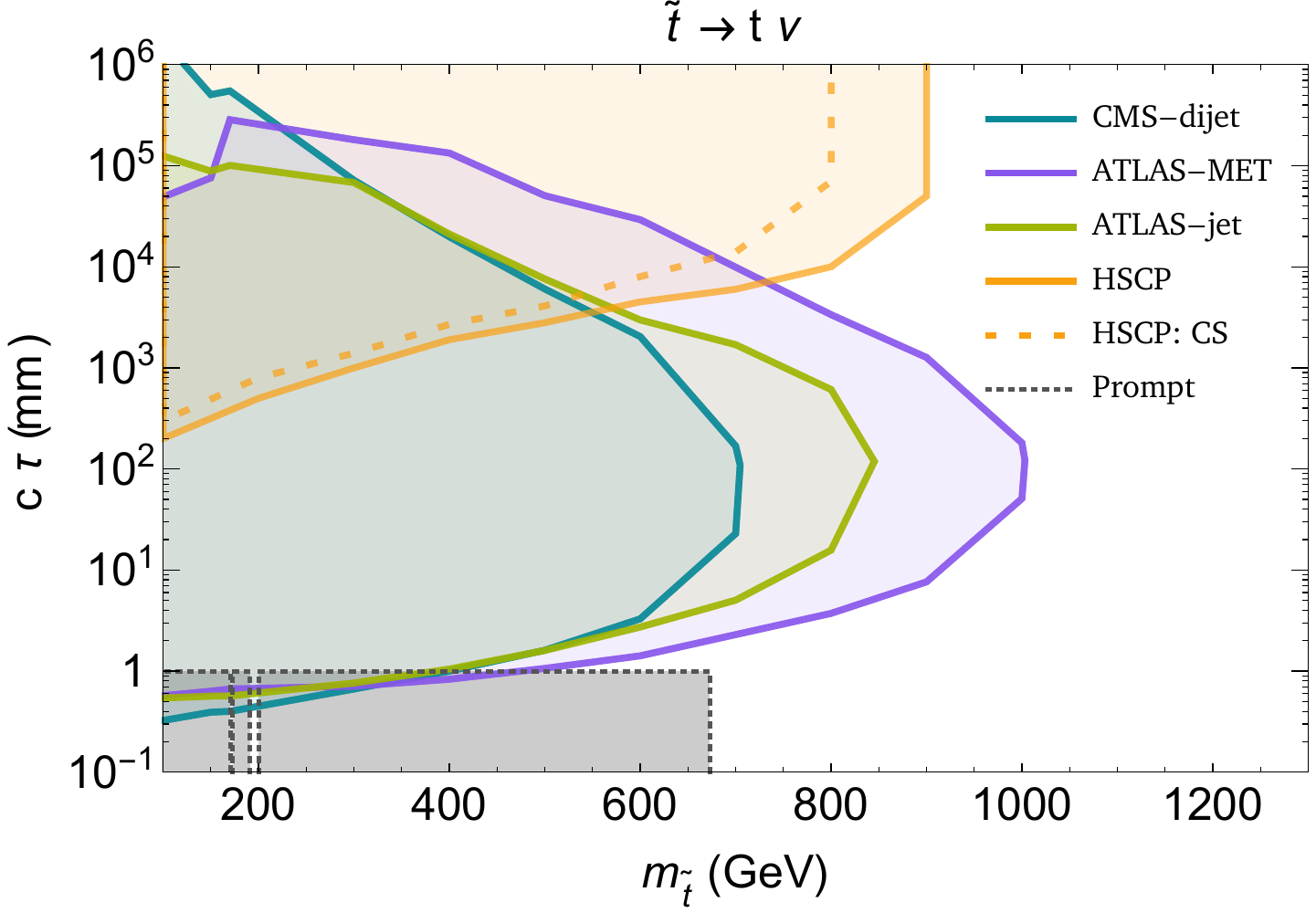}
\includegraphics[width=.48\textwidth]{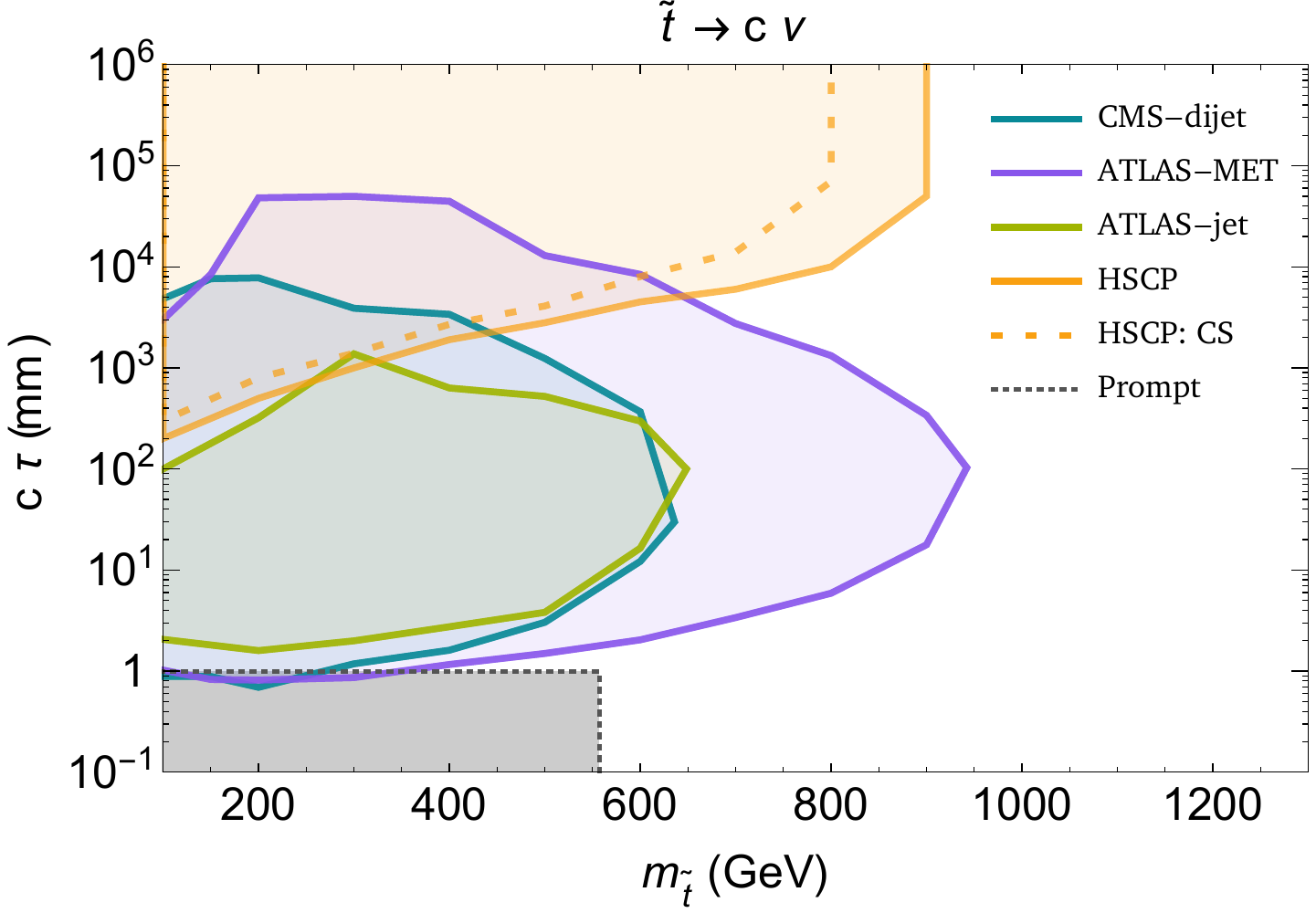}
\caption{95\% CL exclusion curves for stop decays to an up-type quark and a neutrino. In the {\bf left plot} we have $\tilde{t}\to t\nu$ and in the {\bf right plot} $\tilde{t}\to c\nu$.  While all displaced searches of ATLAS~\cite{Aad:2015rba} and CMS~\cite{CMS:2014wda} have been recast, we only display the curves yielding the strongest constraints. Stable searches for HSCPs \cite{Chatrchyan:2013oca} were recast for an unstable R-hadron. The prompt searches here correspond to the R-parity conserving supersymmetric searches~~\cite{Aad:2014mfk,Aad:2014qaa,Aad:2014kra,Chatrchyan:2013xna,Aad:2015gna} \label{stopun}}
\end{figure}

For the case of a stop LSP, we have considered the three final state topologies,  as given in Table~\ref{Tab:summarychannels}. Only direct production of the stop has been simulated and the gluino was assumed to be decoupled for the determination of production cross section. 
Despite the  change in kinematics, we expect the gluino production to mostly affect the overall stop production rate, and therefore the limits presented here can be used to estimate the limits for the non-decoupled gluino case. 

For the $\tilde{t} \to \bar{d}\bar{d}$ final state we have considered all combinations of final state bottom and light 
down quarks. For the $\tilde{t} \to dl^+$ final state, we have considered bottom quarks with all three lepton generations. For the $\tilde{t} \to u\nu$ final state, we studied top and charm quarks.  
For this final state, we have also considered the case where the LSP stop is lighter than the top in which case the stop decays to $W^+ b \nu$ via an off-shell top.

The results are presented in Figs.~\ref{stopdd}-\ref{stopun}. One can see that the whole region of natural stop masses $m_{\tilde{t}}<800$ GeV is excluded, except for purely hadronic prompt decay, $\tilde{t} \to \bar{d}\bar{d}^\prime$. There is also some allowed parameter space by these searches for prompt decays with neutrinos, however we expect that prompt missing energy searches will further constrain some of the 
region.

The bounds derived from the various combinations of bottom and light 
down quarks for the $\tilde{t} \to \bar{d}\bar{d}$ channel are extremely similar.  One would expect the additional displacements of the bottom quarks to decrease the efficiency of vertex reconstruction with respect to the case of two light quarks. However, we found this effect is not highly significant (see Fig \ref{fig:cmseff}).  The reason is that the stop decay vertex contains tracks from the $R-$hadron remnant along with radiation from the two bottom quarks before hadronization. These tracks can be reconstructed as a secondary vertex even if the $B$-meson decays are reconstructed as tertiary vertices.  We expect this to be the case when a bottom quark is exchanged for a down quark for all stop and gluino channels in consideration. Thus, we have confined our simulations for all other channels to include bottom quarks in the final states and have not considered the lighter $\bar{d}$ generations. For the Higgsino, which does not form an $R-$hadron, the bounds for 3rd generation quarks are likely weaker than for light quarks which do not have secondary displacements, but the qualitative differences are expected to be small.

The bounds derived from the $\tilde{t} \to t \nu$ and $\tilde{t} \to c \nu$ final states are also very similar except for the ATLAS DV+jets search.  This difference is due to the fact that the top can decay with larger jet multiplicity than the lighter quarks, thus strengthening the bounds derived from final states with top quarks.  However, since typically, the strongest bounds are derived from the ATLAS DV+MET search for final states with neutrinos, this is not of great significance.  We expect the same to be true for the case of gluino and Higgsino LSPs and therefore consider only tops in the final states for their decays.

The ATLAS DV+MET search bounds are presented only for the final states $b \tau$, 
$t\nu$ 
and $c \nu$, for which this search gives the strongest bounds (ATLAS DV+jets is comparable for the $b \tau$ final state).  The bump in the bound at $\sim 200$ GeV LSP mass is a consequence of the MET $>180$ GeV cut.

\subsection{Gluino LSP}
\begin{figure}[t!]
\center
\includegraphics[width=.48\textwidth]{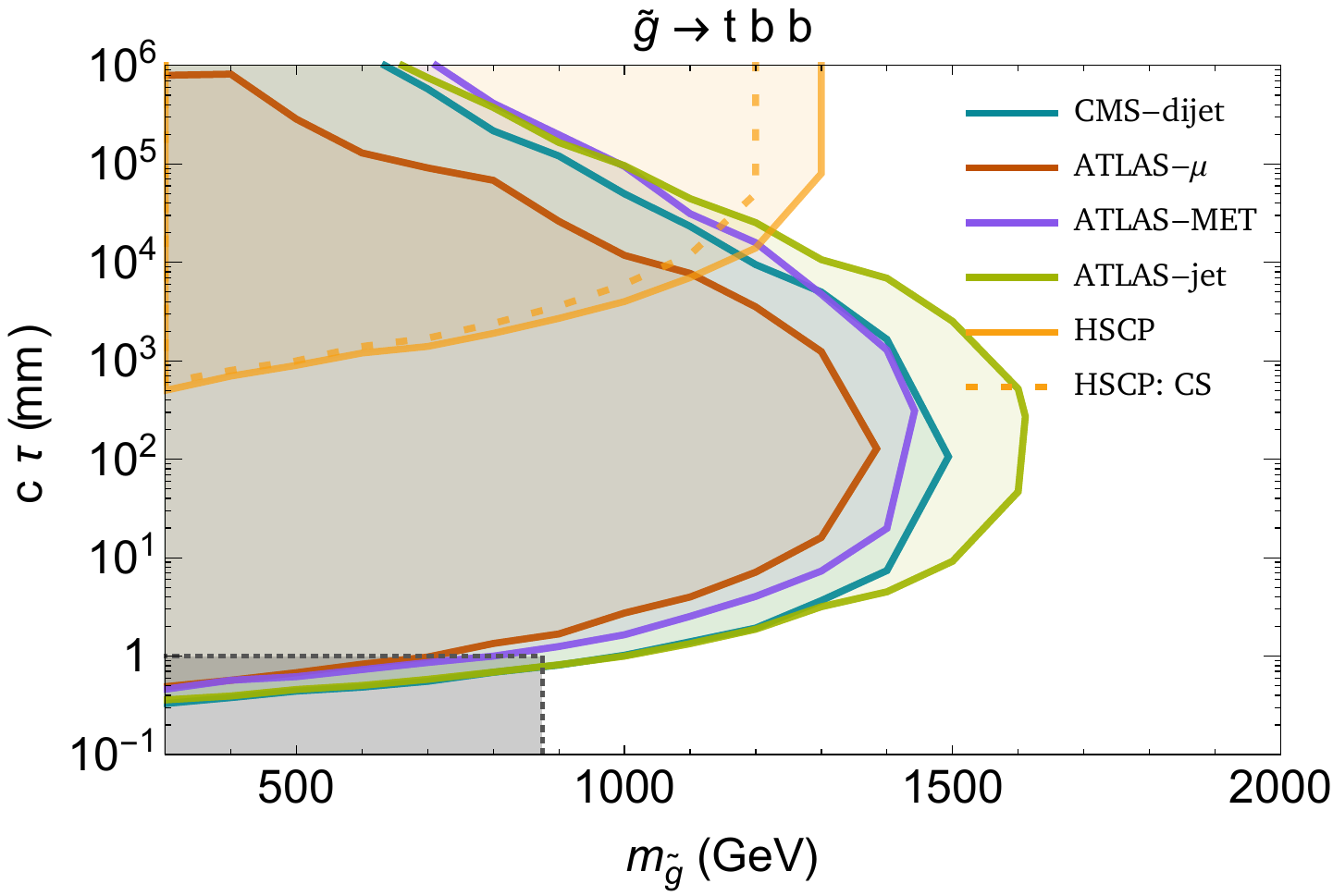}
\includegraphics[width=.48\textwidth]{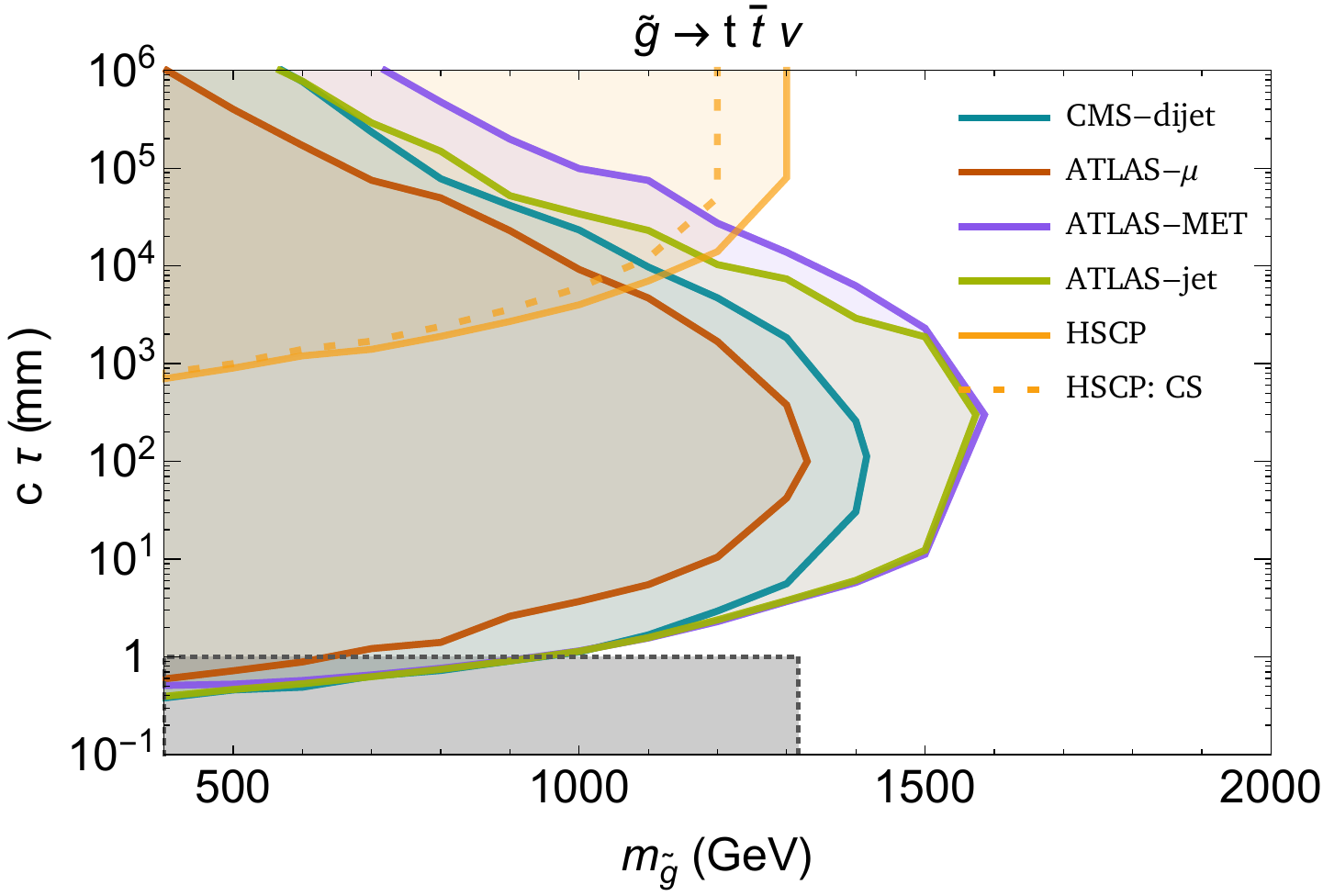}
\caption{95\% CL exclusion curves for gluino decays to $tbb$ ({\bf left}) and $t\bar{t}\nu$ ({\bf right}).  While all displaced searches of ATLAS~\cite{Aad:2015rba} and CMS~\cite{CMS:2014wda} have been recast, we only display the curves yielding the strongest constraints. Stable searches for HSCPs \cite{Chatrchyan:2013oca} were recast for an unstable R-hadron. The prompt searches here correspond to the R-parity conserving supersymmetric searches~\cite{Aad:2014lra,Chatrchyan:2013iqa} and the RPV gluino search~\cite{Aad:2014pda}. \label{gluino1}}
\end{figure}

\begin{figure}[ht!]
\center
\includegraphics[width=.48\textwidth]{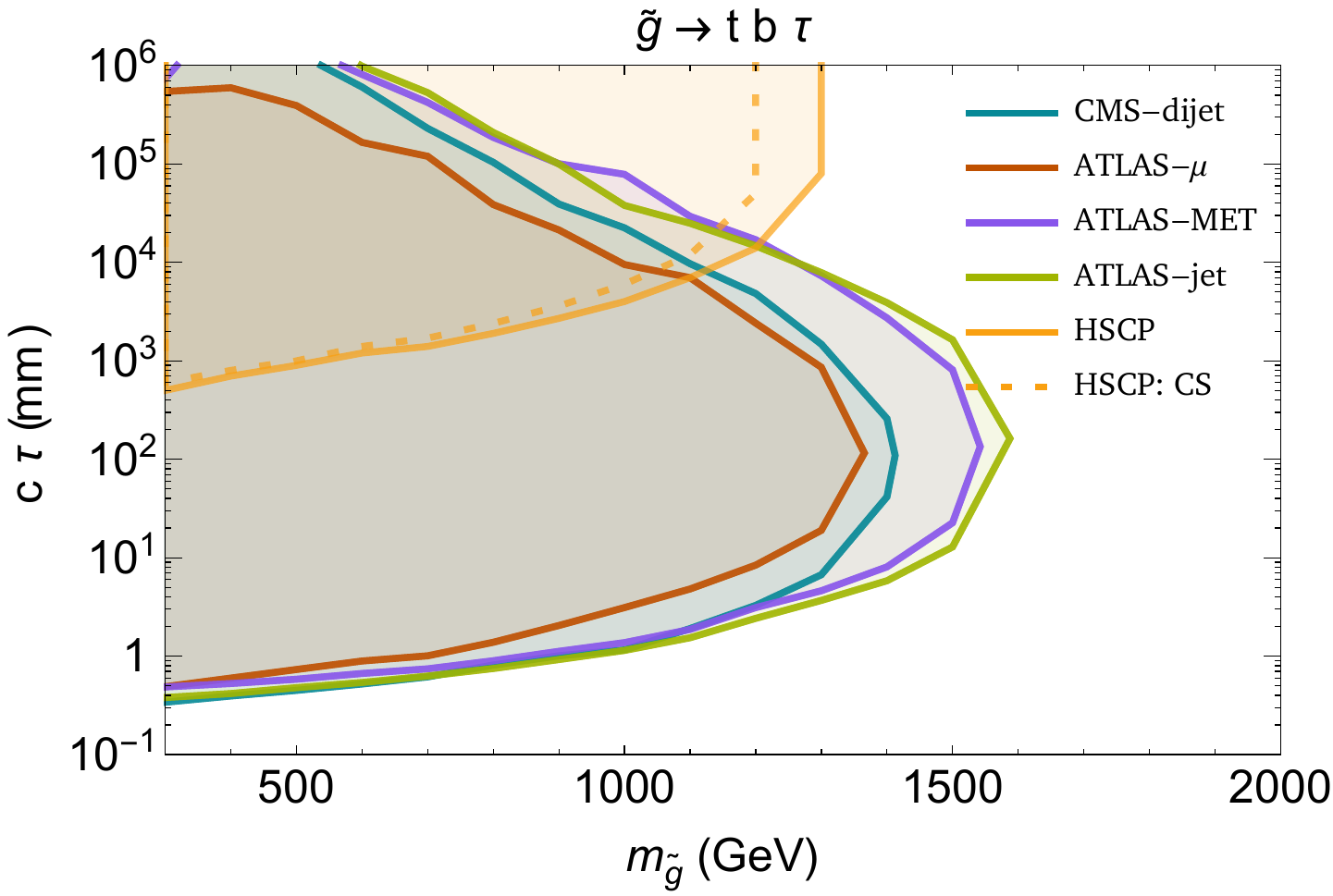}
\includegraphics[width=.48\textwidth]{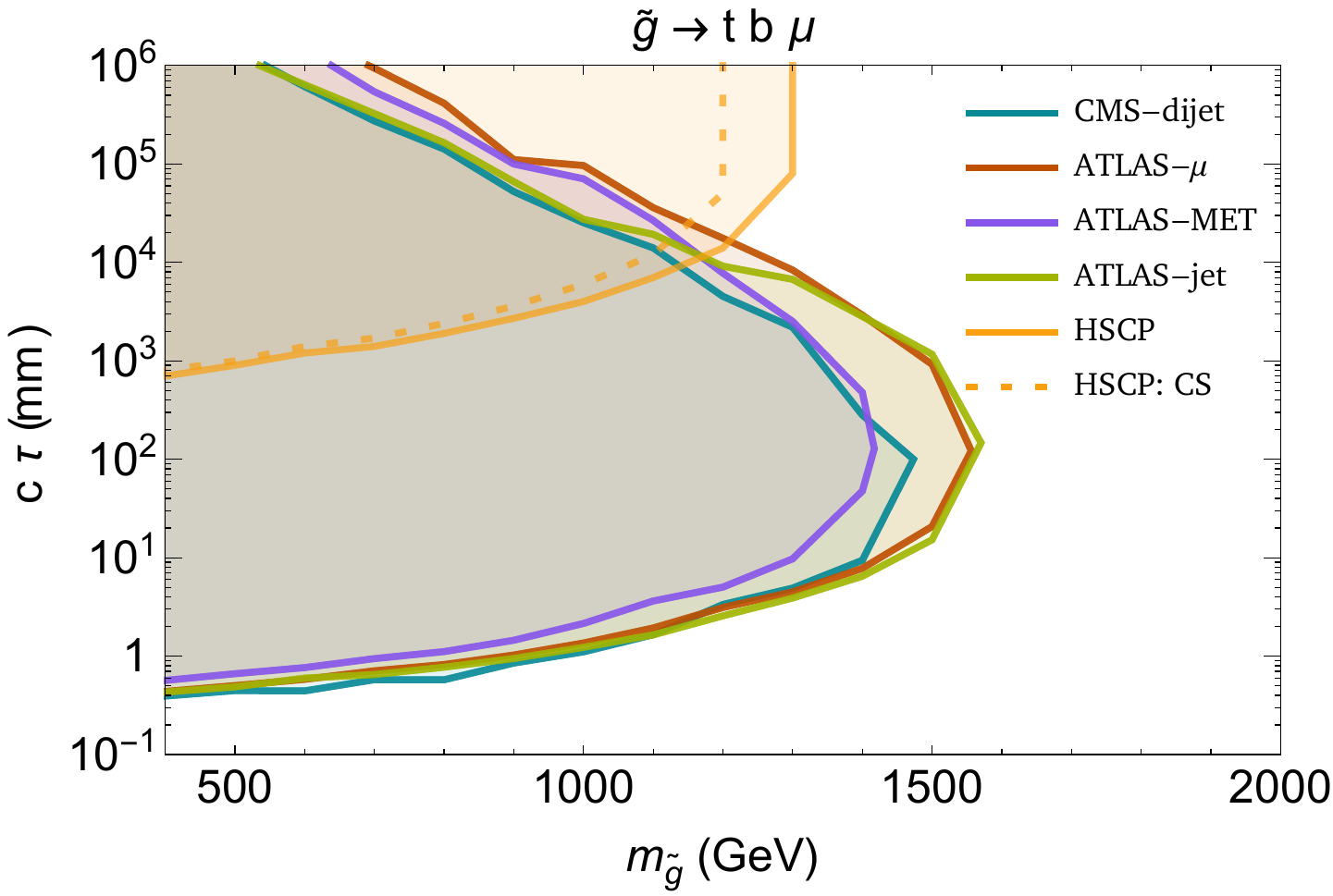}
\includegraphics[width=.48\textwidth]{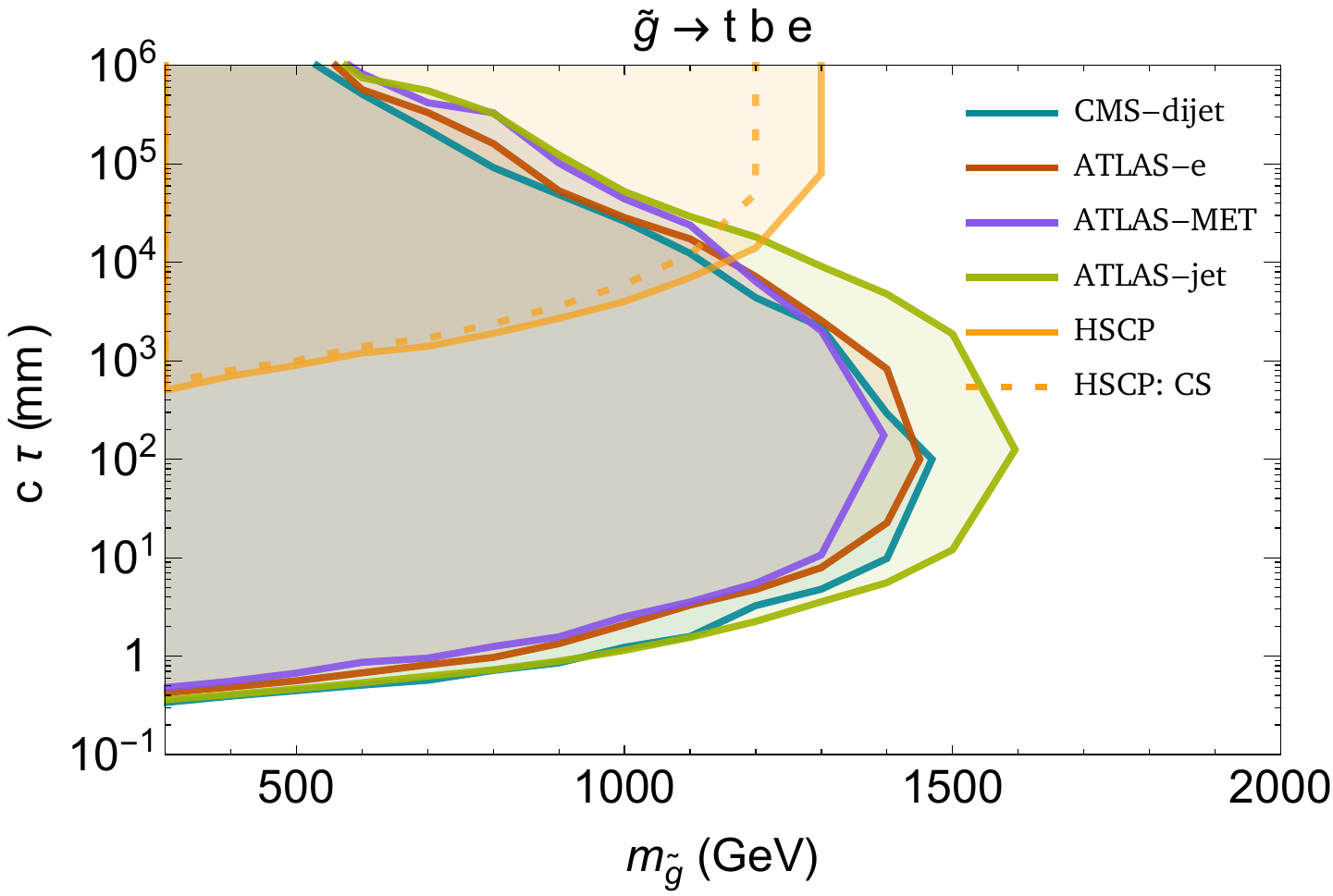}
\caption{95\% CL exclusion curves for gluino decays to a top, a bottom quark and a charged lepton. In the {\bf upper left}  plot we have $\tilde{g}\to tb\tau$, in the {\bf upper right} plot $\tilde{g}\to tb\mu$, while in the {\bf bottom} plot we have $\tilde{g}\to tbe$. Although all displaced searches of  ATLAS~\cite{Aad:2015rba} and CMS~\cite{CMS:2014wda} have been recast, we only display the curves yielding the strongest constraints. Stable searches for HSCPs \cite{Chatrchyan:2013oca} were recast for an unstable R-hadron.  LHC searches for 3 or more leptons, same-sign leptons events, and/or multiple ($b$-)jets are not shown here, and are expected to constrain prompt decays. \label{gluino2}}
\end{figure}

For the case of gluino LSP, we have considered the three types of final states as in Table~\ref{Tab:summarychannels} corresponding to $\tilde{g}\to tbb, t\bar{t}\nu, tbl$.
Only direct production of the gluino has been simulated and the stop has been assumed to be decoupled for the gluino production cross section. The results are presented in Figs.~\ref{gluino1}-\ref{gluino2}. One can see that the whole region of natural gluino masses $m_{\tilde{g}}<1500$ GeV is excluded for long lived gluinos. There may be  a small prompt region for the $tbb$ final state that may still be somewhat natural. 
We expect the $\tilde{g}\to tb\ell$ final states to  be constrained by same-sign dilepton and other relevant prompt searches which we have not attempted to recast here.


As explained above, we have not considered the case of light down quarks in the final state, since the bounds for these are expected to be similar to those with bottom quarks.  For the case of up type quarks in the final state, we only present results for top quarks, i.e. the channel $\tilde{g} \to t \bar{t} \nu$.  As noted above, the decays $\tilde{g} \to t \bar{c} \nu$ and $\tilde{g} \to t \bar{d} \nu$ have very similar constraints for all channels.

\subsection{Higgsino LSP}

\begin{figure}[ht!]
\center
\includegraphics[width=.48\textwidth]{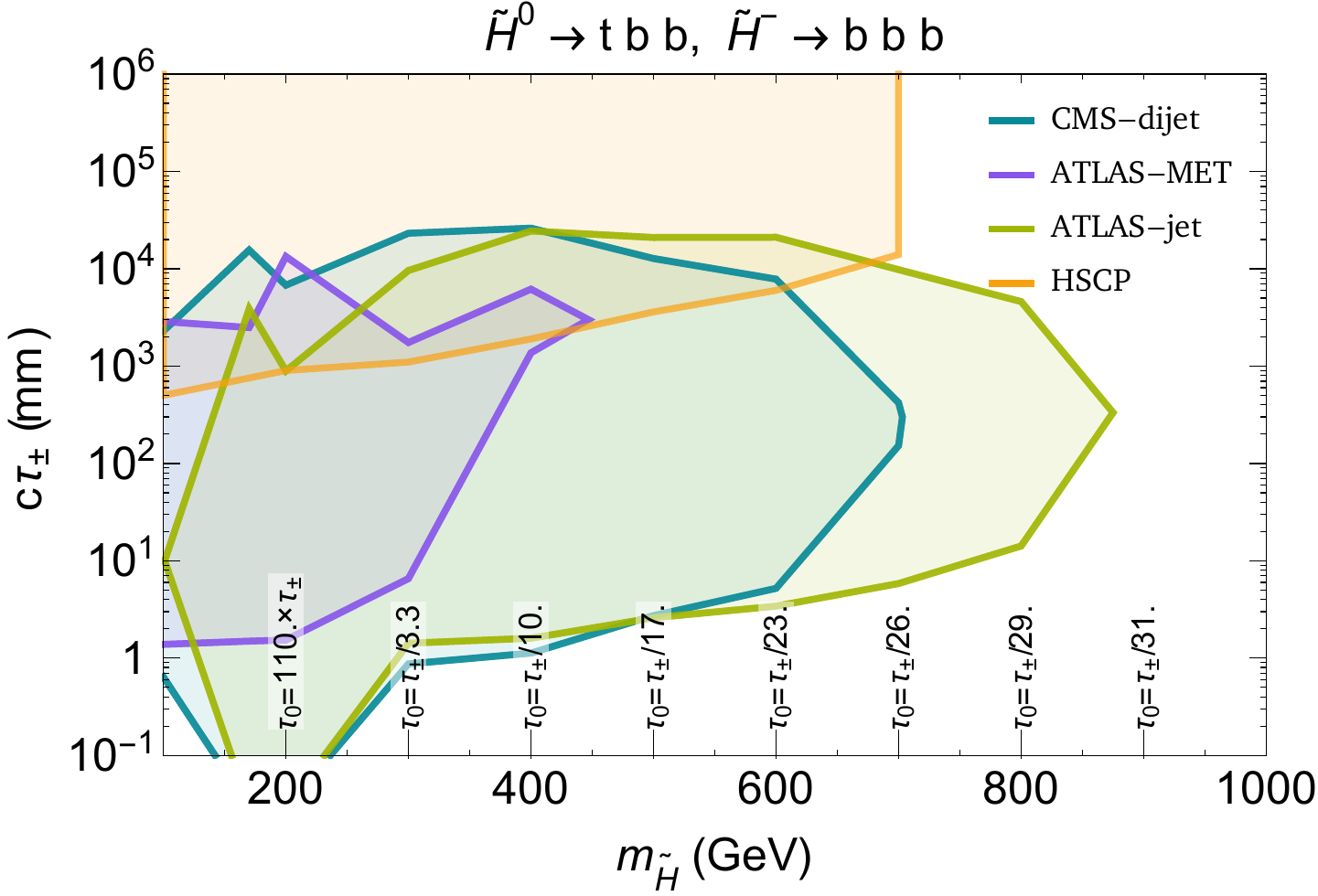}
\includegraphics[width=.48\textwidth]{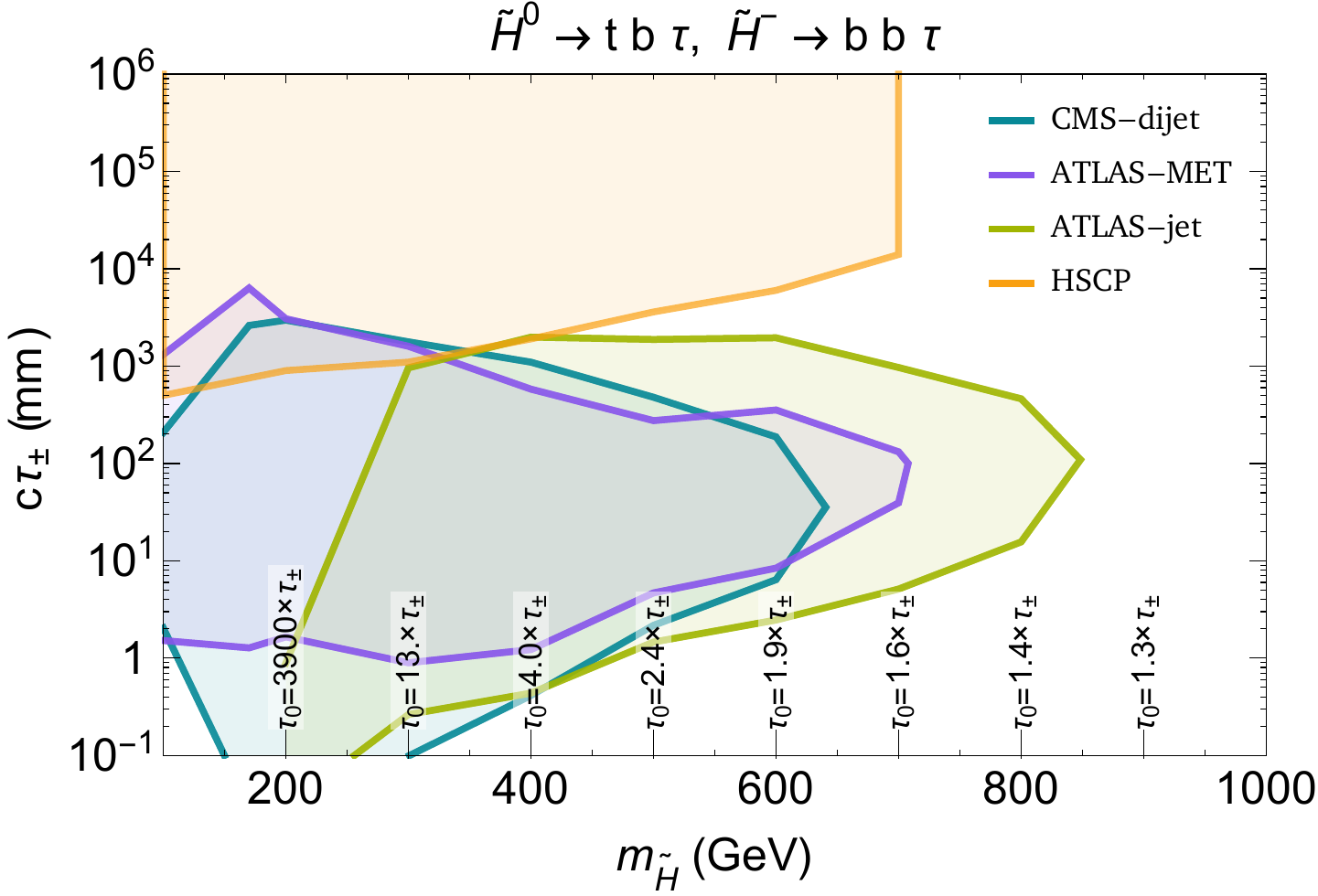}
\includegraphics[width=.48\textwidth]{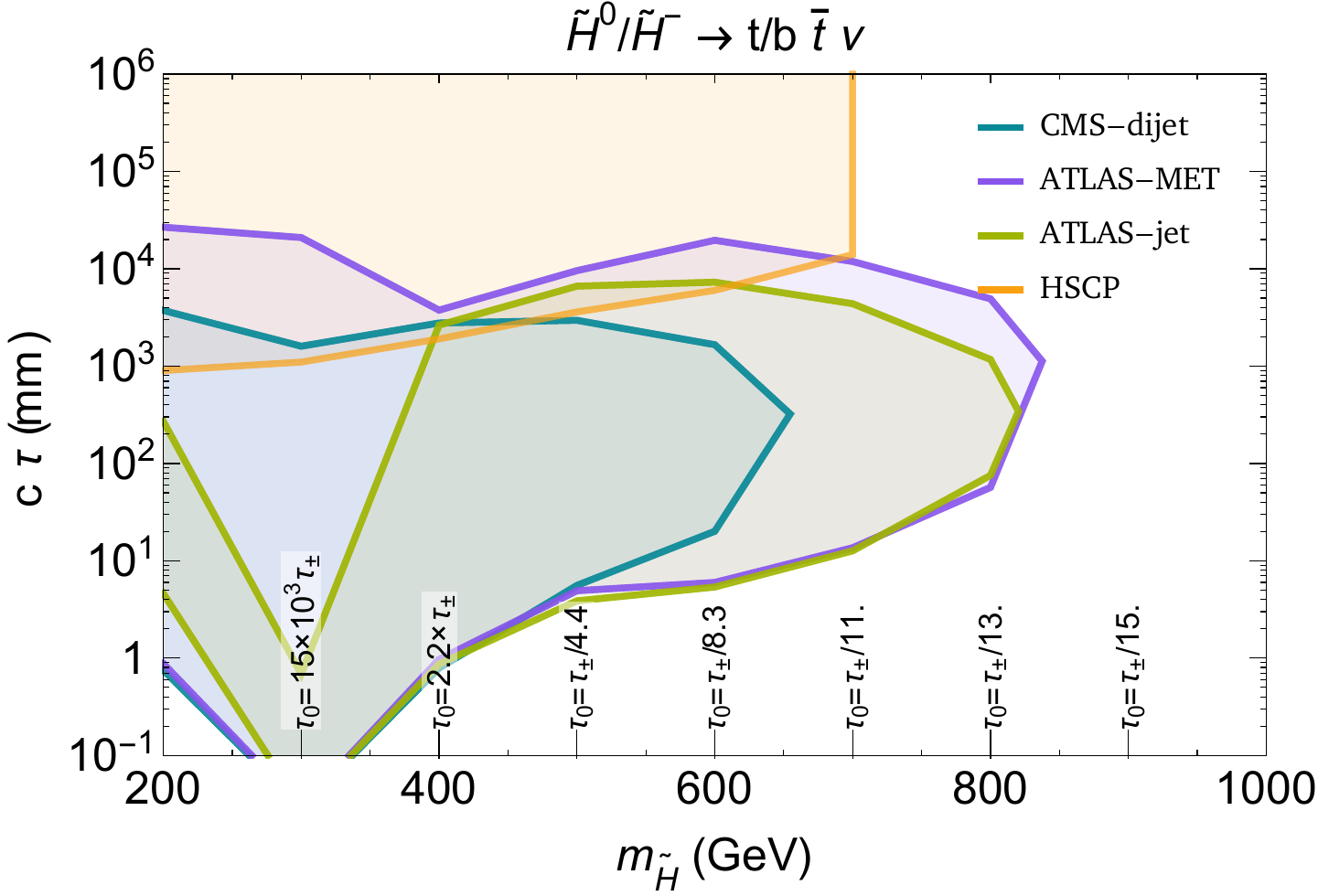}
\caption{95\% CL exclusion curves in the charged Higgino lifetime vs.~mass plane, for Higgsino decays to a top, a bottom quark and a charged lepton. In the {\bf upper left} plot we show  $\tilde{H}^0 \to tbb$ and $\tilde{H}^\mp \to bbb$, in the {\bf upper right} plot $\tilde{H}^0 \to tb\tau$ and $\tilde{H}^\mp \to bb\tau$, and in the {\bf bottom} plot we present $\tilde{H}^0 \to tt\nu$ and $\tilde{H}^\mp \to bt\nu$. The ratio of neutral to charged Higgino lifetime depends on their mass and is indicated on all three plots.  While all displaced searches of  ATLAS~\cite{Aad:2015rba} and CMS~\cite{CMS:2014wda} have been recast, we only display the curves yielding the strongest constraints. Stable searches for HSCPs \cite{Chatrchyan:2013oca} were recast for an unstable Higgsino.  LHC searches for 3 or more leptons, same-sign leptons events, and/or multiple ($b$-)jets are now shown above and are expected to constrain prompt decays. \label{higgsino}}
\end{figure}

For the case of Higgsino LSP, we have considered direct production of charged and neutral Higgsinos. We simulated the production of both particles 
 simultaneously using the corresponding mixing matrices for the relevant value of $\mu \simeq m_{\tilde{H}}$.
Approximately 61\% of produced Higgsinos are charged over the full range of masses, giving HSCP searches sensitivity to long-lived Higgsinos.
The results are plotted in Fig.~\ref{higgsino}. We decouple all the superpartners except for the stops, which we take to be degenerate. This production mechanism may be subdominant to gluino or stop production, which we have not considered. However, we obtain strong bounds from direct production alone, allowing for a conservative limit which rules out Higgsinos up to 800 GeV, except for a region of short lifetimes above $\sim 300$ GeV. This region is expected to be constrained by prompt searches for large jet multiplicity events for $(t/b)bb$ final states, same sign di-lepton for $(t/b)b\tau$ final states and prompt MET searches for $(t/b)t\nu$ final states as well as other prompt searches.  Recasting these searches is beyond the scope of this paper.

We have simulated three channels, given in  Table~\ref{Tab:summarychannels}. For each coupling under consideration, the charged and neutral Higgsinos decay differently.  For the $\eta^{\prime\prime}$ coupling, the decay is $\tilde{H}^0 \to t b b$ and $\tilde{H}^\mp \to b b b$.  For the $\eta^{\prime}$ coupling, the decay is either $\tilde{H}^0 \to t b \tau$ and $\tilde{H}^\mp \to b b \tau$ or $\tilde{H}^0 \to t t \nu$ and $\tilde{H}^\mp \to b t \nu$.  In each case, the neutral (charged) Higgsino decays into an off-shell stop and a top (bottom) quark.  The stop then decays via the RPV coupling.  For down type quarks in the final state, we consider only bottoms quarks since the bounds for lighter generations are very similar.  For up type quarks in the final state we consider only top quarks, again since the bounds for lighter generations are expected to be almost the same.  Furthermore, we consider only $\tau$ leptons in the final states since we expect the bounds for lighter generations to differ very slightly as was shown for the case of a gluino LSP.

The charged and neutral Higgsinos have different lifetimes due to differences in phase space of the final states. The plotted lifetime is the $\tilde{H}^\mp$ lifetime. For $\tilde{H}^0$ below 200 GeV, the Higgsino decays via an off shell top. At 100 GeV, the neutral Higgsino is effectively stable due to phase space suppression. The qualitatively different features of these results are due to the presence of two different average lifetimes in each data sample. For example, in Fig.~\ref{higgsino}, the ATLAS DV+MET search becomes important for masses where the neutral Higgsino becomes long-lived and contributes to MET. Prompt lifetimes are ruled out at 200 GeV due to the presence of longer lived neutral Higgsino which decay within the tracker.

\section{Conclusions\label{sec:Conclusions}}

We have considered the experimental bounds on supersymmetric theories with R-parity violation and a long-lived LSP. The main searches providing the constraints are the CMS displaced dijet search~\cite{CMS:2014wda} and the ATLAS search for multitrack displaced vertices + $\mu/e$/jets/MET~\cite{Aad:2015rba}, which we have fully recast and applied all cuts and vertex reconstruction procedures. In addition we also considered the CMS HSCP search~\cite{Chatrchyan:2013oca}, as well as some of the leading prompt searches whose exclusion bounds can be directly applied without recasting. Our main results are the exclusion plots in the $m_{LSP}-\tau_{LSP}$ plane in Figs.~\ref{stopdd}-\ref{higgsino}, where the LSP is a stop in Figs.~\ref{stopdd}-\ref{stopun}, a gluino in Figs.~\ref{gluino1}-\ref{gluino2}, or a Higgsino in Fig.~\ref{higgsino}. The various plots for a given LSP correspond to different decay channels either from the ordinary holomorphic RPV operator of (\ref{eq:WRPV}) or the non-holomorphic dRPV operators of (\ref{eq:KRPV}). All regions with significant displacement, $c\tau \gtrsim 1$ mm, are excluded for the natural SUSY mass regions  $m_{\tilde{t}} \lesssim 800$ GeV, $m_{\tilde{g}} \lesssim 1500$ GeV or $m_{\tilde{H}} \lesssim 800$ GeV. The only significant unconstrained regions are those corresponding to prompt stop decays to dijets or prompt Higgsino decays.

\section*{Acknowledgements}
We are grateful to Avner Sofer and Nimrod Taiblum for many important clarifications regarding~\cite{Aad:2015rba}.
We also thank Jeff Dror, Gauthier Durieux, Yuri Gershtein, Kevin McDermott, Wee Hao Ng, Torbjorn Sjostrand, Peter Skands, and Jordan Tucker  for useful discussions.  
C.C., E.K. and S.L. are supported in part by the NSF grant PHY-1316222. O.S. and T.V. are supported in part by a grant from the Israel Science Foundation.  T.V. is further supported by  the US-Israel Binational Science Foundation, the EU-FP7 Marie Curie, CIG fellowship and by the I-CORE Program of the Planning Budgeting Committee and the Israel Science Foundation (grant NO 1937/12).

\appendix
\section*{Appendix}
\section{Data and Simulations\label{sec:Data}}

\subsection{Monte Carlo Tools}

Parton level events are generated using the {\tt Madgraph5} event-generator~\cite{Alwall:2011uj}, utilizing model files created by {\tt Feynrules}~\cite{Alloul:2013bka}.
Hadronization, parton showering, and $R$-hadron physics  are simulated using {\tt Pythia~8} \cite{Sjostrand:2014zea}, which allows the $\epsilon_{\alpha\beta\gamma}$ color flow associated with the $q q\bar{d}^*$ and $\bar{u} \bar{d} \bar{d}$ operators.  Long-lived particles are displaced before passing events from {\tt Madgraph} to {\tt Pythia}.
For the detector simulation, we use {\tt Delphes~3} \cite{deFavereau:2013fsa} and {\tt ROOT~6.02} \cite{Brun:1997pa} with the default ATLAS and CMS detector geometries, resolutions, and efficiencies (excluding tracking efficiency). The {\tt Delphes~3} detector has an inner detector (ID), electromagnetic and hadron calorimeters (ECAL/HCAL), and a muon system (MS). Charged SM particles are propagated in a uniform magnetic field in the tracker. {\tt FastJet}~\cite{Cacciari:2011ma} is used to cluster jets according to track and calorimeter information from the Delphes particle-flow reconstruction. Tracking vertex resolution is taken to be perfect, but nearby vertices are merged in the vertex reconstruction procedures.

\subsection{Tracking Efficiency}

The standard tracking simulation provided in {\tt Delphes~3}, which assumes tracking efficiencies for particles originating from the PV, provides a poor reproduction of the actual tracking efficiency for tracks originating from displaced vertices. The track reconstruction efficiencies in ATLAS~\cite{ATLAS-CONF-2013-092} and CMS \cite{CMStalk1,CMStalk2} drop off rapidly for vertices displaced from the origin.  For the ATLAS and CMS displaced vertex searches presented in this work, offline tracking techniques (``re-tracking'') are used to regain some of the lost efficiency. Even with displaced tracking algorithms, the probability to reconstruct a track in the CMS or ATLAS inner detectors depends on the track's orientation and the displacement of its production vertex. This effect is incorporated in our analysis by assigning displaced tracks efficiencies based on simulated displaced track reconstruction efficiencies of the ATLAS and CMS detectors. For the ATLAS detector, we use an efficiency parametrized by the transverse impact parameter, $d_0$, and transverse momentum, $p_T$, of each track. 
In addition to tracking efficiency falloff with $d_0$, ATLAS vertex reconstruction efficiency also drops off steeply at a radial distance corresponding to the second pixel barrel layers of the ID. To account for this effect, we add an additional factor of 75\% to the track reconstruction efficiency for tracks which originate outside of the second pixel layer. The CMS track reconstruction efficiencies are simulated under Run 1 tracking conditions \cite{CMStalk1} and are parametrized as a function of production radius. 
A cut on track transverse impact parameter $d_0 < 30$ cm is imposed, motivated by displaced tracking efficiencies for cosmic muons \cite{CMStalk2}. The final reconstruction efficiencies do not strongly depend on the the shape and scale of the tracking efficiency nor on their parameterization either as a function of production radius or impact parameter. 

Tracks of charged $R$-hadrons which decay in the tracker are assumed to not reconstruct and are neglected in this study. It is possible that some of these tracks could be reconstructed by ATLAS or CMS, but this depends on the tracker hits registered in the detector before the charged $R$-hadron decays and the details of the tracking algorithm. Tracks of charged $R$-hadrons which decay in the beam pipe do not register hits in the detector and therefore are not reconstructed. 
$R$-hadrons decaying in the tracker may have tracks which exhibit multi-prong decays 
or which 
 disappear altogether, and may also 
not reconstruct properly depending on the reconstruction algorithm. 
The track from the charged $R-$hadron would appear to be prompt because it points back to the primary vertex (PV).
The ATLAS multitrack DV search ignores prompt tracks with transverse impact parameter $d_0 < 2$ mm in reconstructing displaced vertices.
The CMS dijet search allows for only one prompt track to be associated to each displaced jet, so the search could retain sensitivity to events in which the charged $R$-hadron tracks are
reconstructed. Therefore, it is reasonable to neglect the track of the $R$-hadrons in this study. The curvature of charged $R$-hadron trajectories, typically small due to the large momentum, is neglected.

\section{Searches}
\label{sec:DetailedSearches}

\subsection{ATLAS DV+$\mu / e /$jets/MET }

In this section, we describe the ATLAS~\cite{Aad:2015rba} multitrack search for displaced vertices along with our procedure. Trigger and selection cuts vary between the different final state topologies while displaced vertex reconstruction remains the same for all cases. 

\subsubsection{Trigger Requirements}

For the {\bf DV+muon} search, a trigger muon is required to have:
\begin{itemize}
\item{$p_T>50\text{ GeV}$}
\item{pseudo-rapidity $|\eta| < 1.07$ to be within the MS acceptance}
\item{tracker hits both in the ID and MS}
\end{itemize}
For the {\bf DV+electron} search, a photon trigger is implemented since electrons with large transverse impact parameters may not have a measured track in the inner detector and may be identified as photons.  The search triggers on either one or two photons with large transverse momentum.  The requirement is:
\begin{itemize}
\item{either one photon with  $p_T>120 \text{ GeV}$ or two photons, each with $p_T>40 \text{ GeV}$}
\end{itemize}
For the {\bf DV+jets} search, jets are reconstructed using the anti-$k_t$ algorithm with size parameter $\Delta R = 0.6$. The trigger requires:
\begin{itemize}
\item{four jets with $p_T>80\text{ GeV}$  or five jets with $p_T>55\text{ GeV}$ or six jets with $p_T>45 \text{ GeV}$}
\end{itemize}
The trigger requirement for the {\bf DV+MET} search is:
\begin{itemize}
\item $E_T^{\text{miss}}>80 \text{ GeV}$
\end{itemize}

\subsubsection{Event Selection}

Offline cuts filter displaced tracks which undergo a re-tracking procedure prior to final analysis.  The details of these cuts and the re-tracking are given in \cite{Aad:2015rba}.  Following this procedure, final event selection cuts are implemented for the various channels.  
In our analysis, we require the final event selection criteria on the reconstructed leptons, jets, and MET since they are more restrictive than the trigger requirements in all cases. The trigger leptons must have a truth level origin within the ID.

All events must have a reconstructed DV with track multiplicity $N_{tr}\geq5$ and invariant mass $m_{DV}>10 \text{ GeV}$, where each track is assigned a charged pion mass for the invariant mass determination.\\
For the {\bf DV+muon} search, the cuts are similar to the muon trigger, requiring:
\begin{itemize}
\item{$p_T>55\text{ GeV}$}
\item{$|\eta|<1.07$}
\end{itemize}
The {\bf DV+electron} search cuts require the electron candidate to have:
\begin{itemize}
\item{$p_T>125 \text{ GeV}$}
\item{$|\eta|<2.48$}
\end{itemize}
Furthermore, for both lepton searches, the lepton track must be displaced, corresponding to a transverse impact parameter which satisfies $d_0>1.5 \text{ mm}$ and is required to pass within $0.5 \text{ mm}$ of a reconstructed DV.\\
Cuts for the {\bf DV+jets} search are stricter than the trigger requirements requiring:
\begin{itemize}
\item{four jets with $p_T>90\text{ GeV}$ or five jets with $p_T>65\text{ GeV}$ or six jets with $p_T>55 \text{ GeV}$}
\end{itemize}
For the {\bf DV+MET} search, the requirement is:
\begin{itemize}
\item $E_T^{\text{miss}} > 180 \text{ GeV}$
\end{itemize}

There is a subtlety regarding whether LSPs which escape the tracker contribute to the transverse energy of the event for the {\bf DV+MET} search.  A charged $R$-hadron or Higgsino which decays outside of the entire detector may be reconstructed as a muon from its tracks in the MS and therefore be included in the energy of the event. LSPs, charged or neutral, which decay outside of the ID but within the calorimeter or within the MS may contribute to the energy of the event either from calorimeter deposits or from the tracks of its decay products.
In our analysis, we treat such LSPs as follows:  Charged LSPs which traverse the entire detector and leave a track in the ID are reconstructed as muons and the transverse momentum of the LSP's track does not contribute to MET.  Transverse energy deposited by the decay products of either charged or neutral LSPs which decay within the calorimeter is removed from the MET of the event with the exception of neutrinos in the final state. For LSPs decaying in the MS, charged decay products are not included in the MET of the event. We neglect the calorimeter deposits from the $R$-hadrons themselves since they are expected to be small. This MET reconstruction procedure for decays in the various regions of the detector is a conservative approach since, in reality, some of this energy is not reconstructed and will contribute to the MET of these events.

\subsubsection{Vertex Reconstruction}

Events satisfying the trigger requirements must also contain at least one reconstructed DV. Vertex reconstruction in the ATLAS search is performed considering only displaced tracks with $d_0>2\text{ mm}$ and $p_T>1\text{ GeV}$.  All tracks are extrapolated in the direction opposite to their momentum.  Tracks which pass the criteria are used to reconstruct DVs by means of an algorithm detailed in \cite{Aad:2015rba} which clusters seed vertices iteratively.  A seed vertex is defined as two coinciding tracks which each satisfy the criteria $\vec{d} \cdot \hat{p}>-20 \text{ mm}$, which ensures consistency between the position of the vertex and momentum of the track.  Here $\vec{d}$ is defined as the distance vector between the position of the DV and that of the first primary vertex and $\vec{p}$ is the seed track's momentum.  After the tracks are clustered into vertices, any reconstructed DVs within 1 mm are merged into a single vertex. 

After vertex reconstruction, DVs are further required to have transverse distance $L_{xy}<300 \text{ mm}$ and $|z_{DV}|< 300\text{ mm}$ with respect to the PV.  In order to minimize background from tracks originating at a PV, each DV must also have a transverse distance of at least 4 mm from any primary vertex. Finally, DVs situated within dense regions of the detector are vetoed using a 3D map of the transverse plane of the detector~\cite{Aad:2015rba}.  

For our analysis, vertex reconstruction is identical for all the final state searches.  Tracks considered for vertex reconstruction are required to have $p_T>1 \text{ GeV}$ and impact parameter $d_0>2$.  The truth level origins of all tracks must also satisfy $4\text{ mm}<L_{xy}<300 \text{ mm}$ and $|z|<300\text{ mm}$. Our vertex reconstruction algorithm searches through all tracks that satisfy these criteria and iteratively clusters their origins. First, two track origins which are within a 1 mm radius from one another are grouped together as a DV.  The DV position is defined as the average position of all of the track origins in the group. DVs are then combined if their distances are less than 1 mm apart.
Furthermore, at least two tracks in each DV are required to satisfy $\vec{d} \cdot \hat{p}>-20 \text{ mm}$ in consistency with the seed vertex track requirement detailed above. DVs are vetoed at positions in the transverse plane which are mapped by the ATLAS study~\cite{Aad:2015rba} as dense material regions of the detector. 

For the DV+lepton channels, lepton tracks are extrapolated in the opposite direction of their momenta.  The extrapolated track is then required to have a maximum distance of closest approach to the reconstructed DV of 0.5 mm.
This procedure has important implications for signals with final state bottom or charm quarks since these decays result in additional displacement from the LSP decay vertex, and the DV clustering algorithm could, in principle, reconstruct these vertices separately.  The ATLAS study associates each track with only one DV. However, the lepton track is not required to fit to the reconstructed DV and may therefore pass within 0.5 mm from multiple vertices, as in the case of the decay products of very boosted bottom or charm quarks.  This is taken into account in our procedure since truth level tracks are always associated with only one DV and lepton tracks are extrapolated and not required to be directly associated with the DV. 

\subsubsection{Efficiencies}

\begin{figure}[t!]
\center
\includegraphics[width=.48\textwidth]{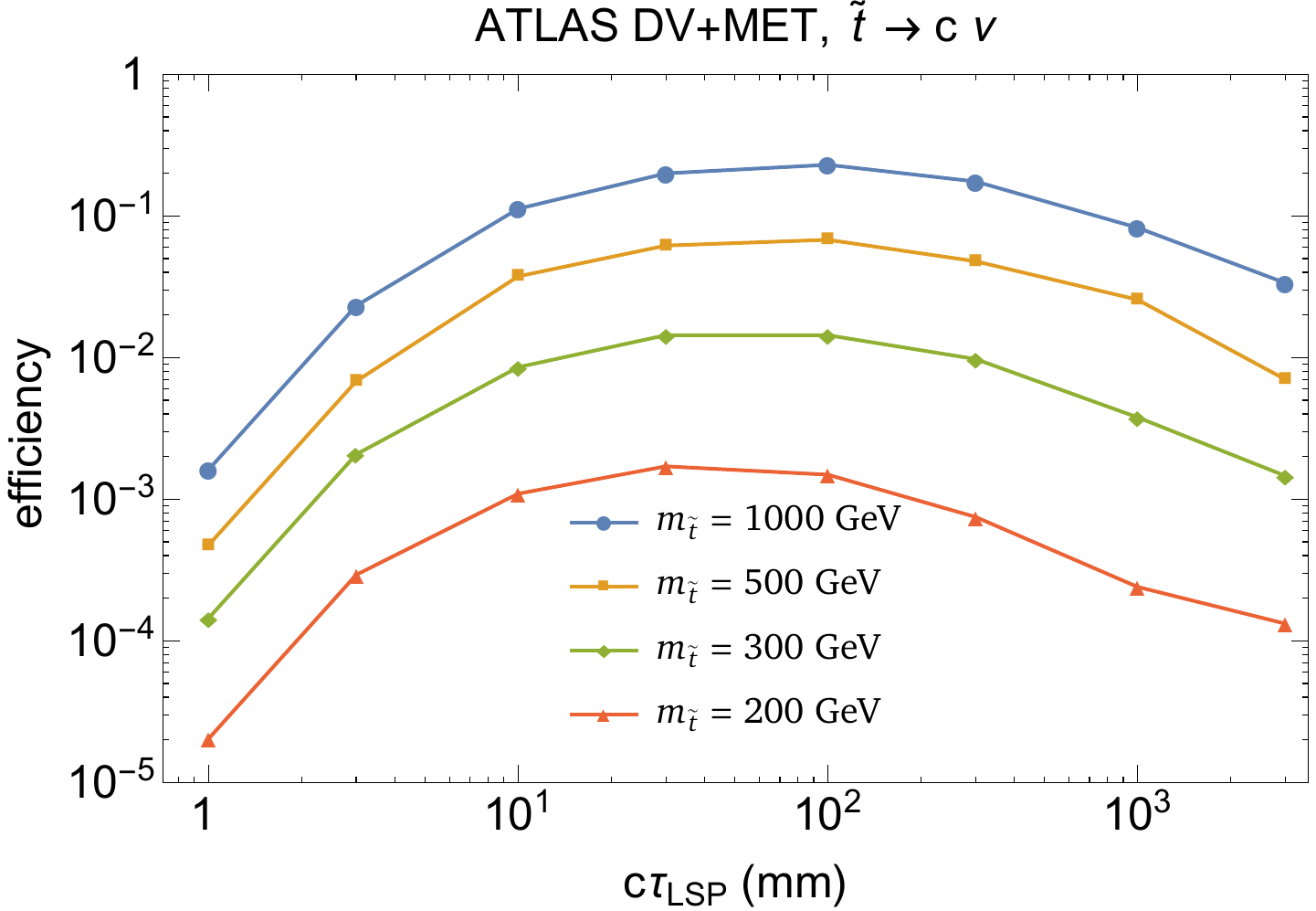}~
\includegraphics[width=.48\textwidth]{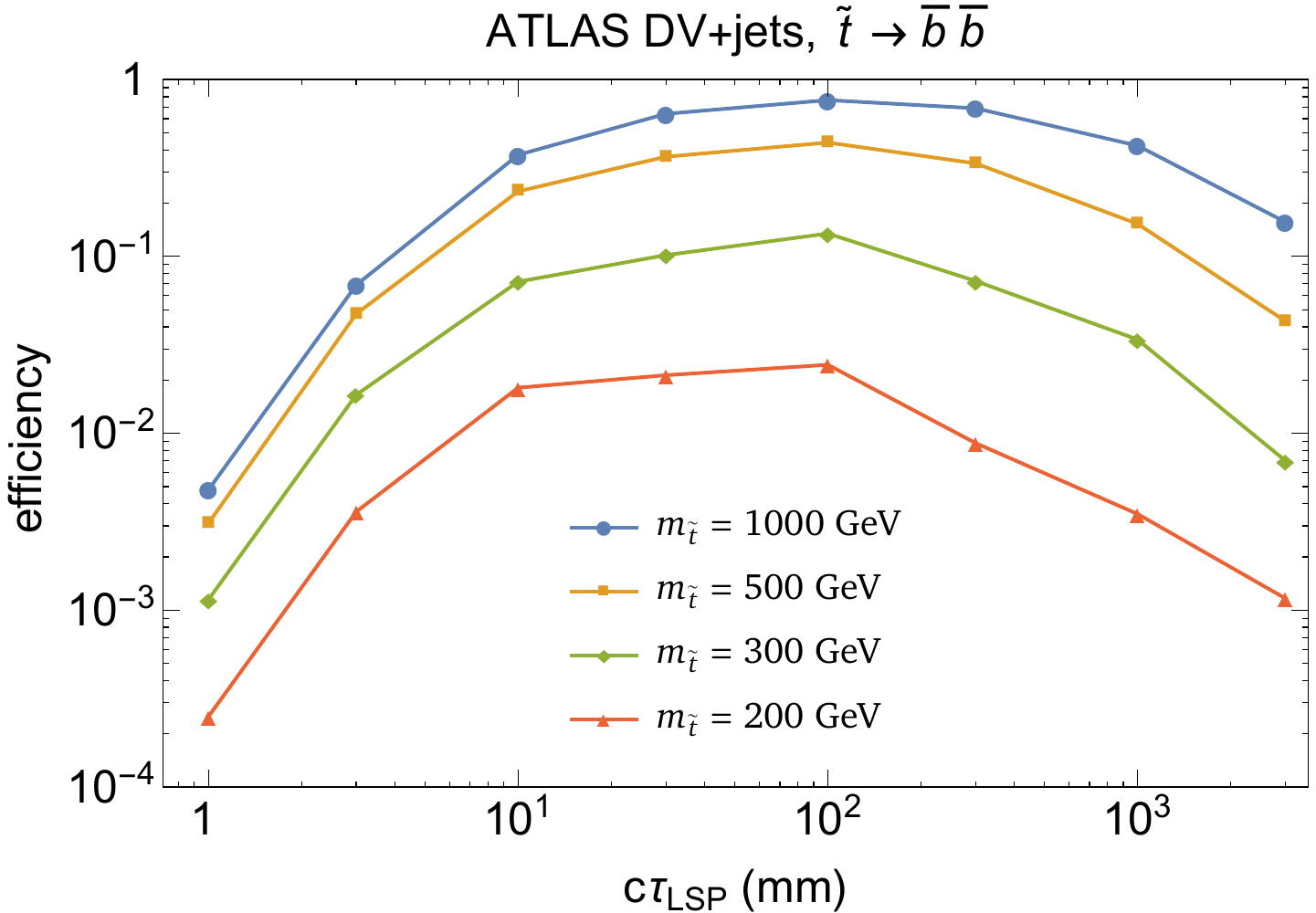}\\
~\\

\includegraphics[width=.48\textwidth]{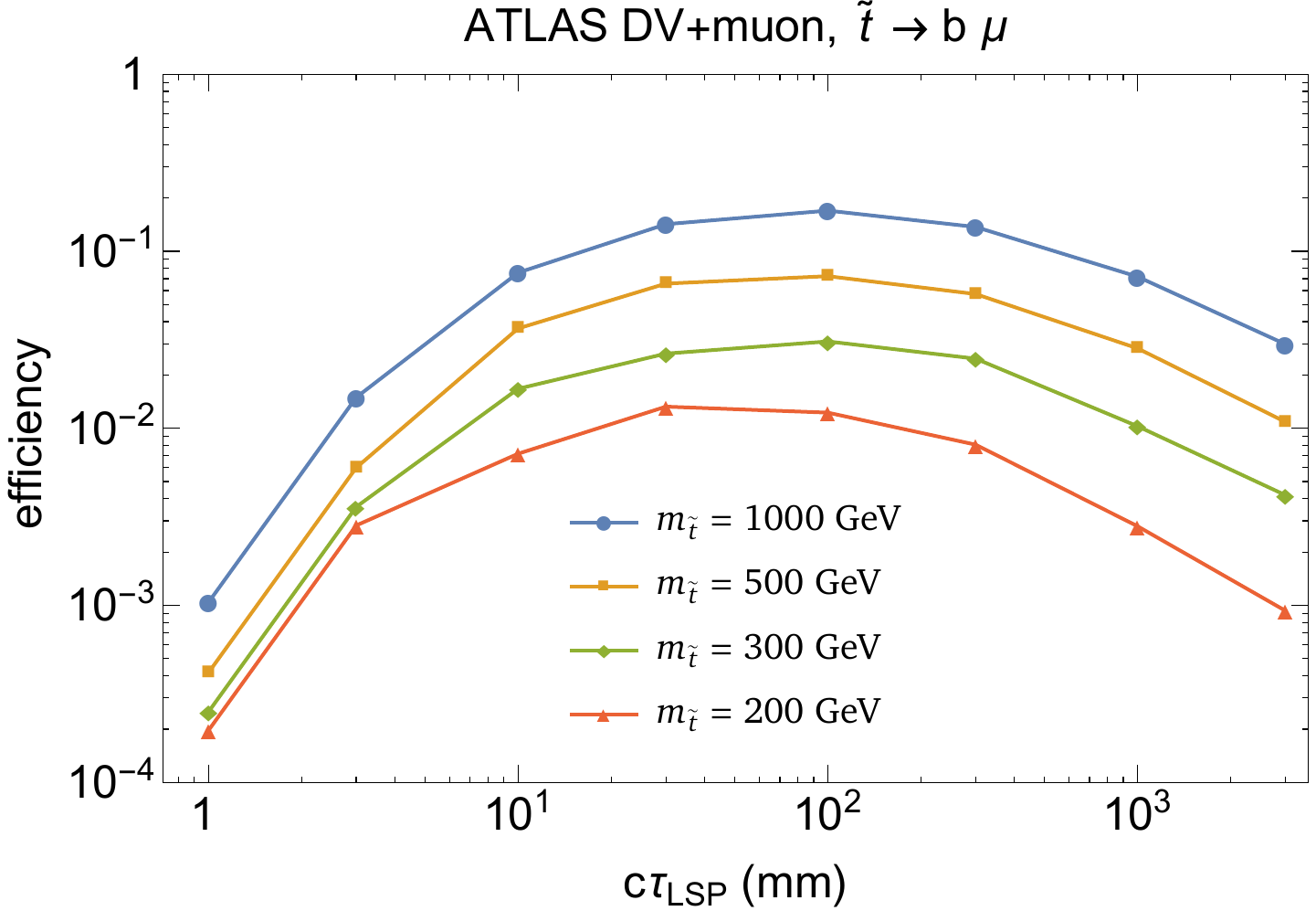}~
\includegraphics[width=.48\textwidth]{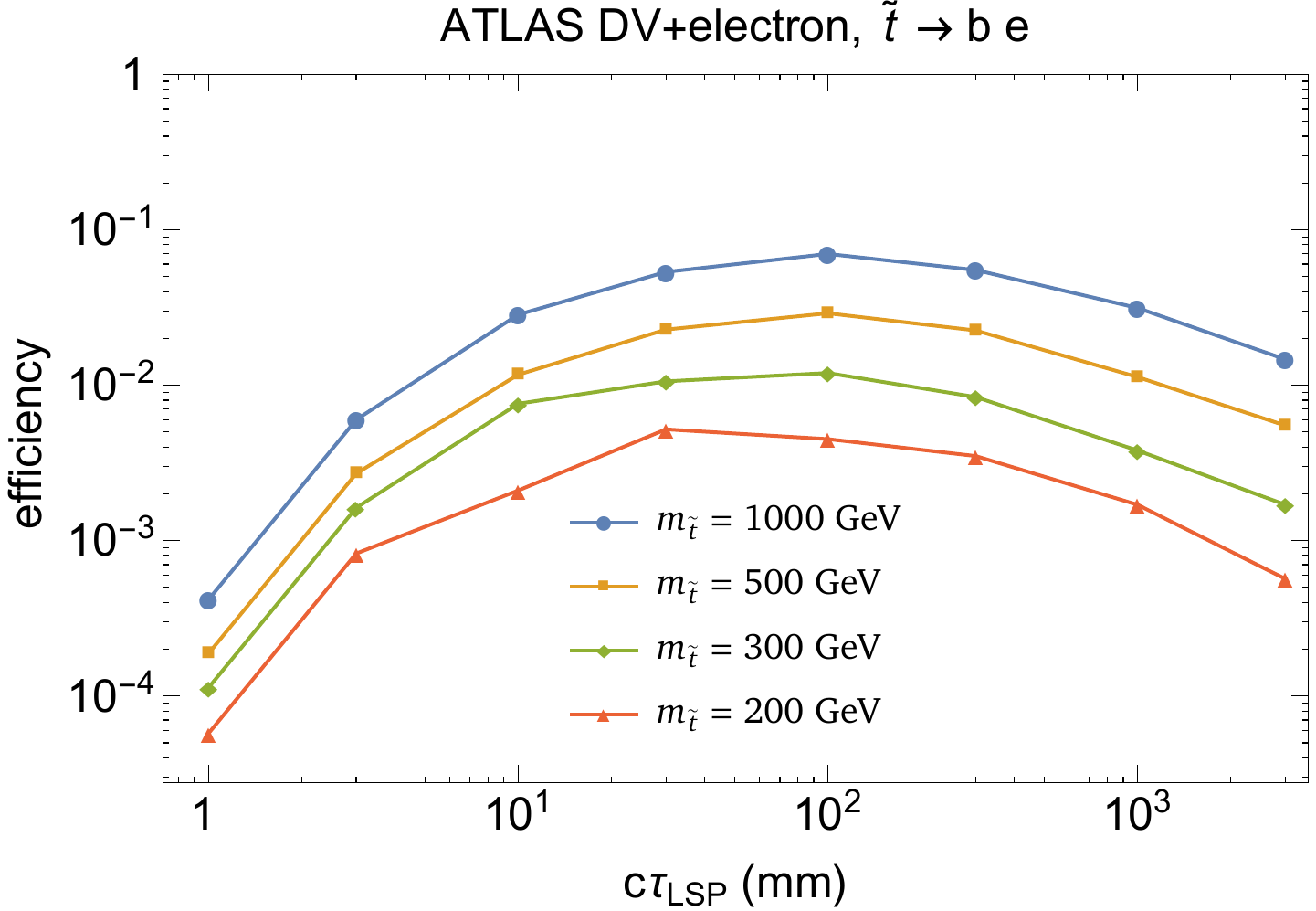}
\caption{Example event-level efficiencies for the ATLAS displaced vertex searches. Efficiencies are shown for $m_{\tilde{t}} =$ 200 (red), 300 (green), 500 (orange), 1000 (blue) GeV. Efficiencies generally increase with increases stop mass. \label{fig:atlaseff}}
\end{figure}

With these assumptions for vertexing and tracking efficiency, we reproduce the lifetime dependence of the efficiencies for the models and channels provided in \cite{Aad:2015rba}. For the {\bf DV+jets} and {\bf DV+MET} channels, only event-level efficiencies are calculated since the cuts for these channels are required at the event level and not at the vertex level. For the {\bf DV+lepton} channels, vertex-level efficiencies are calculated and then translated into event-level efficiencies. We accept up to two DVs per event and calculate a DV level efficiency, $\epsilon_{DV}$.  In consistency with the ATLAS study, for the DV+lepton channels, the event level efficiency, $\epsilon_{ev}$, is defined as the probability for a signal event containing two DVs to satisfy all selection criteria for at least one of the DVs in the event.  This definition relates to the vertex level efficiency, $\epsilon_{DV}$, through,
\begin{equation}
\epsilon_{ev} = 2 \epsilon_{DV} - \epsilon_{DV}^2.
\label{event_level_eff}
\end{equation}
Example event level efficiencies are shown in Fig.~\ref{fig:atlaseff}  and have been used in our calculation of the exclusion plots in Sec.~\ref{sec:Results}.

\subsection{CMS Displaced Dijet}

CMS~\cite{CMS:2014wda} performs two searches for signals with high and low average transverse displacements. We recast the study using the high-$\langle L_{xy} \rangle$ selection criteria, which is strong enough to place limits on the full range of displaced lifetimes for the models considered in this paper.

\subsubsection{Trigger Requirements and Offline Cuts}

CMS uses a dedicated displaced-jet trigger for this search which requires
$H_T > 300$ GeV, where $H_T$ is the sum of the transverse energy of all the jets in the event with $p_T > 40 $ GeV and $|\eta| <3$, and at least two displaced jet candidates (isolated leptons are also considered) each satisfying: 
\begin{itemize}
	\item $p_T > $ 60 GeV and  $|\eta| < 2$,
	\item at most 2 associated tracks with three-dimensional impact parameter less than 300~$\mu$m,
	\item at most $15\%$ of the jet energy is carried by associated tracks with transverse impact parameters less than 500 $\mu$m.
\end{itemize}
At trigger level, jets are clustered according to calorimeter information only.\\

The offline event selection cuts are similar to the triggers. The total transverse energy requirement is raised to $H_T > 325$ GeV for the fully reconstructed jets, using anti-$k_T$ algorithm with $\Delta R = 0.5$. 
 For our jet reconstruction, we use track and calorimeter information from the Delphes particle-flow reconstruction, and cuts are imposed according to the offline selection criteria made by CMS.

\subsubsection{Dijet Reconstruction and Final Event Selection}

Every jet pair is checked for consistency with the displaced dijet hypothesis. 
Displaced tracks, defined by transverse impact parameter $d_0 > 0.5$ mm, associated to each jet pair are grouped together and fitted to a common secondary vertex. Tracks originating at a radial distance greater than 50 cm have zero acceptance because the strip tracker layers no longer contain 3D information, and the search requires track seeds to be within the part of the tracker with stereo tracking information. The full analysis uses an adaptive vertex fitter which is not easily implemented. We mimic the vertex fitter by merging the origin of two tracks if their truth-level origins are within a distance of 1 mm. If more than one displaced vertex is found between the two jets, we take the one with largest track multiplicity. Each of the two jets must have at least one track originating from the secondary vertex, which must be significantly displaced from the primary vertex of the event. The search considers the transverse displacement of a secondary vertex, $L_{xy}$, to be significant if $L_{xy} > 8 \ \sigma_{L_{xy}} $, 
where $\sigma_{L_{xy}} $ is the uncertainty in $L_{xy}$.
The primary vertex resolution for CMS is 12 microns in each dimension \cite{Chatrchyan:2014fea}, which is negligible compared to the uncertainty in the secondary vertex. We take the uncertainty in $L_{xy}$ be a constant 300 $\mu$m for all DVs as a conservative choice, requiring the displaced vertex to be at a distance of 2.4 mm away from the primary vertex in the transverse plane. For decays with short decay lengths (for which this cut is relevant), the resolution of $L_{xy}$ is better than 300 $\mu$m given the good resolution near the center of the detector.
This cut does not significantly affect the final efficiencies because decays which occur this close to the PV begin to have tracks which point back to the PV and fail the requirements of the displaced-jet trigger.

For final event selection, further selection criteria are imposed for each displaced dijet candidate passing the above criteria. 
Clusters of maximal track multiplicity are formed from displaced tracks associated to dijet candidates based on the transverse displacements of each individual track, $L_{xy}^{track}$.
$L_{xy}^{track}$ is determined by finding the intersection of the dijet transverse momentum (the sum of the two reconstructed jet $p_T$) with the trajectories of the individual tracks. In the full analysis, the $L_{xy}^{track}$ of individual tracks are clustered using an algorithm with size parameter $0.15 \ L_{xy}$. For this recast, we approximate this procedure by grouping tracks with $L_{xy}^{track}$ within a transverse distance $0.15 \ L_{xy}$ together in a cluster.
The associated track cluster must also satisfy:
\begin{itemize}
\item at least one track from each jet belongs to the cluster
\item invariant mass of cluster $>$ 4 GeV
\item sum of track $p_T$ $>$ 8 GeV
\end{itemize}
Tracks are assigned the mass of the charged pion for the purpose of determining the invariant mass.

Finally, a background discriminant is formed by the secondary vertex track multiplicity, the cluster track multiplicity, the cluster RMS of $L_{xy}^{track}$, and the fraction of secondary vertex tracks having positive signed impact parameter. We neglect the RMS and signed impact parameter discriminants since we do not implement the clustering algorithm used by the search and the vertex and cluster track multiplicity affects the value of the full discriminant most significantly
because background DVs
typically have low track multiplicities compared to signal DVs. We use the distributions given in \cite{CMS:2014wda} to form the vertex and cluster multiplicity discriminant.
The final event selection for the high $\langle L_{xy} \rangle$ search requires the background discriminant to be less than 0.8, prompt track energy fraction for each jet  less than $ 9 \%$, and no more than one prompt track associated to each jet.

\subsubsection{Efficiencies}

\begin{figure}[t!]
\center
\includegraphics[width=.48\textwidth]{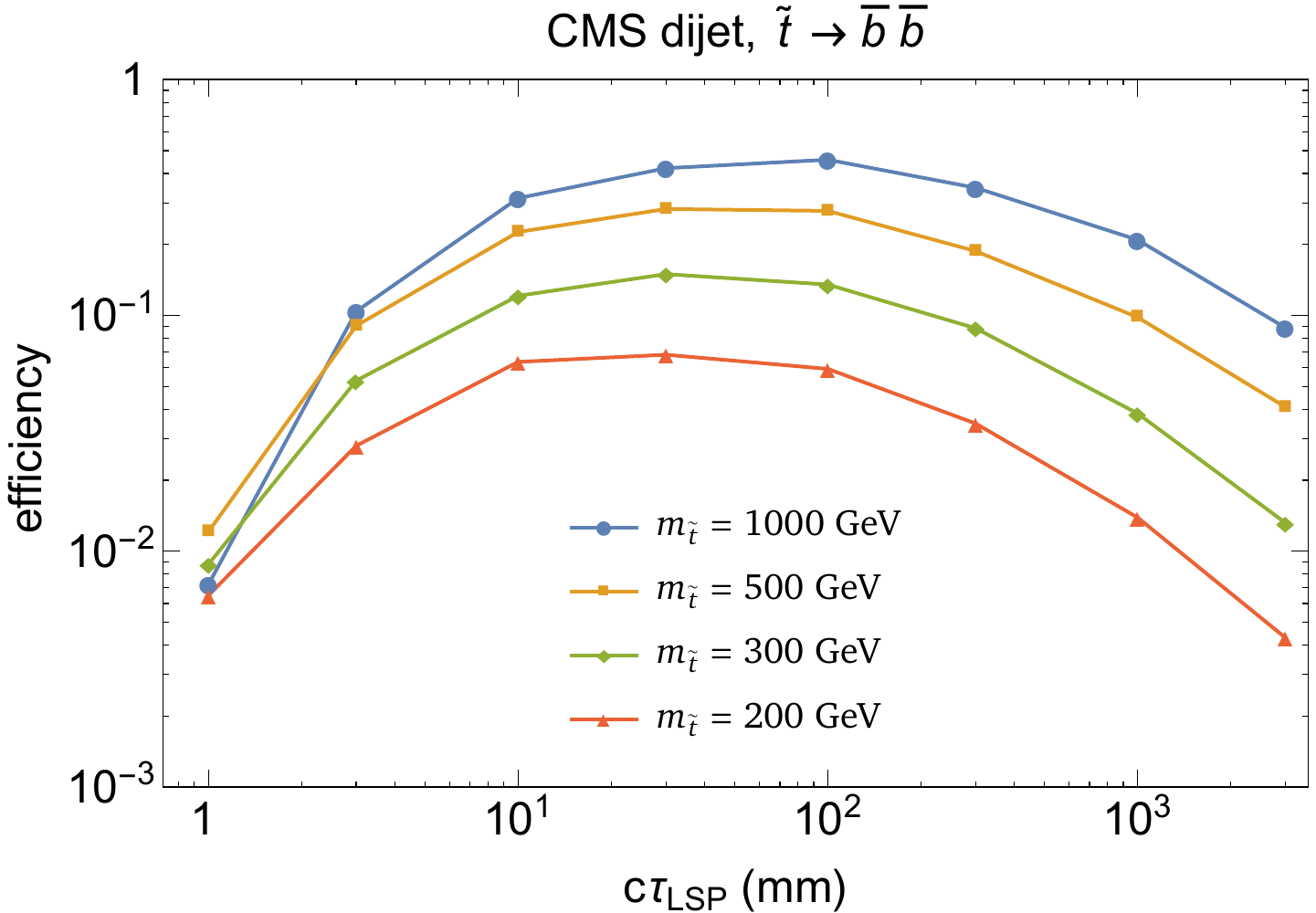}~
\includegraphics[width=.48\textwidth]{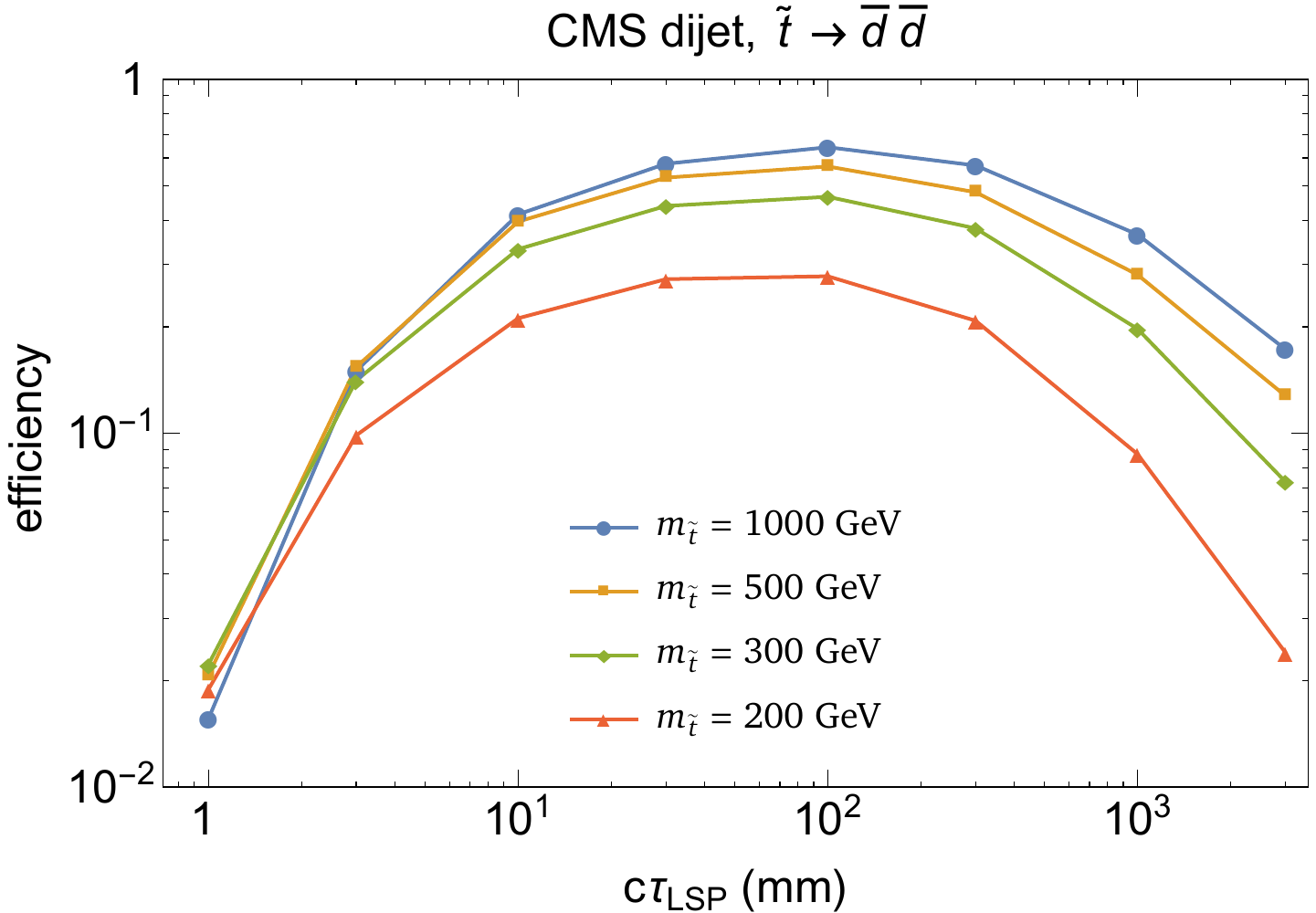}
\caption{Example event-level efficiencies for the CMS displaced dijet search. Efficiencies are shown for $m_{\tilde{t}} =$ 200 (red), 300 (green), 500 (orange), 1000 (blue) GeV. Efficiencies generally increase with increases stop mass. Due to the additional displacement of the $b$-jets, the efficiencies are lower in the  ${\tilde{t}} \to \bar{b} \bar{d}$ decay channel than in decays to $d$- and $s$-quarks. \label{fig:cmseff}}
\end{figure}

The high $\langle L_{xy} \rangle$ search observed 1 data event with an expected background of $1.13 \pm 0.15 \pm 0.50$. This places the excluded number of signal events for $95 \%$ confidence level at 3.7. The results are interpreted in the context of two models:  (1) a heavy scalar particle, $H_0$, decaying into a pair of long-lived neutral $X_0$ particles, which then decay into dijets, and (2) squark pair production, in which each squark decays via $\tilde{q} \to q (\tilde{\chi}_1^0 \to  u \bar{d} \mu)$.  The study considers $X_0$ and $\tilde{\chi}_1^0$ proper lifetimes from 0.1 to 200 cm. We match the reported efficiencies results  within 20\% over the full range of lifetimes for both 2-body and 3-body final states.

Example event level efficiencies are shown in Fig.~\ref{fig:cmseff}  and have been used in our calculation of the exclusion plots in Sec.~\ref{sec:Results}.

\subsection{CMS Heavy Stable Charged Particles}
If the LSP lifetime is sufficiently long, searches for heavy stable charged particles (HSCP) may be sensitive to the scenarios considered in this paper.  Here we briefly describe the CMS search presented in~\cite{Chatrchyan:2013oca}.

Triggered events must have either a reconstructed muon with $p_T > 40$ GeV measured in the ID or a large missing transverse energy $E_T^{miss} > 150$ GeV. The L1 muon trigger can accept slowly moving particles arriving in the MS within the 25 ns of the proton bunch crossing or within the following 25 ns time window before the next bunch crossing. Neutral $R$-hadrons or HSCPs that become neutral do not leave tracks in the muon system. 
Consequently, they are rejected by the online particle flow algorithm and are not reconstructed as objects. However, since it only leaves 10-20 GeV of energy in the calorimeter, there is large large $E_T^{miss}$  in the event.  In such cases, the HSCPs are still identified by the anomalously high energy loss in the inner tracker. The offline data selection requires a track with $p_T > 45$ GeV, $|\eta| < 2.1$, and a $dE / dx > 3$ MeV/cm. 

It is straightforward to reinterpret the search results for unstable $R$-hadrons. The excluded cross sections derived from the full CMS analysis already includes the acceptance and reconstruction efficiencies for stable $R$-Hadrons. The upper limits for stop and gluino production cross section are rescaled by the percentage of events passing the $p_T$ and $|\eta|$ cut that survive the length of the detector. The cut on $dE / dx$ is neglected due to our limited detector simulation. 
Efficiencies for the colored particle to decay outside the detector is found by simulating parton-level events for stop production with masses between 100 GeV - 1 TeV and gluino production with masses between 100 GeV - 1.5 TeV. The detector is taken to be a cylinder with radius of 7m and 11m half-length~\cite{CMS:2014qwa}, and the long-lived particle is required to survive the entire detector distance. This is a conservative choice as it neglects stops that decay in the muon tracker which may still be picked up by the full analysis due to the presence of high $p_T$ tracks in the MS. 

The results are interpreted in terms of stable gluinos and stops, placing upper limits of 1322  (1233) GeV and 935 (818) GeV  in the cloud  (charge-suppressed) model, respectively. CMS performs a ``tracker-only,'' ``tracker+time-of-flight,'' and ``muon-only'' search. The ``tracker-only'' search is sensitive to standard $R$-hadrons and the charge-suppressed scenario because it does not rely on a charged $R$-hadron tracks in the muon system, so we choose to recast this search for the case of unstable $R$-hadrons. For charged higgsinos, we apply the ``tracker+time-of-flight''
search.

\subsection{Prompt Searches}

For our analysis, prompt searches have not been recast.  Instead we have studied a partial subset of prompt searches whose bounds can be directly applied to the scenarios of interest. Recasting other searches may possibly result with more competitive limits than the ones considered here:
\begin{itemize}
\item $\tilde{t} \to \bar{d}\bar{d}$.  The main prompt searches are the paired dijet searches performed at CMS~\cite{Khachatryan:2014lpa}. Constraints for light quark searches are taken to be valid for $b$-quark final states.

\item $\tilde{t} \to d \ell^+$.   We consider leptoquark searches from the Tevatron and the LHC. Searches for leptoquarks are performed for "generation" leptoquarks, i.e., searches are performed for $d e^+$ \cite{Acosta:2005ge,Abazov:2009ab,CMS:2014qpa}, $s \mu^+$ \cite{Abulencia:2005ua,Abazov:2008np,CMS:zva}, and $b \tau^+$ \cite{Aaltonen:2007rb,Abazov:2010wq,Khachatryan:2014ura} decays of the leptoquark. Constraints for light quark searches are taken to be valid for $b$-quark final states. 

\item $\tilde{t} \to c \bar{\nu}$.   We consider ATLAS supersymmetry search for charm squarks, $\tilde{c} \to c \tilde{N}$, where $\tilde{N}$ is the LSP \cite{Aad:2015gna}. The applicable bounds are those for $m_{\tilde{N}} = 0$~GeV. These bounds are much stronger than the monojet search~\cite{Aad:2014nra}.

\item $\tilde{g} \to t \bar{t} \nu$.  We study LHC supersymmetry searches for $\tilde{g} \to t \bar{t} \tilde{N}$. The applicable bounds are those for $m_{\tilde{N}} = 0$~GeV. The strongest bounds are derived from the 0-1 leptons and ($\ge 3$) $b$-jets  search from  ATLAS~\cite{Aad:2014lra} and 1 lepton and ($\ge 6$) jets search  from CMS~\cite{Chatrchyan:2013iqa}.

\item $\tilde{t} \to t \bar{\nu}$.  We consider LHC supersymmetry searches for $\tilde{t} \to t \tilde{N}$, where $\tilde{N}$ is the LSP. The applicable bounds are those for $m_{\tilde{N}} = 0$~GeV. The strongest bounds are derived from the 1-lepton searches from  CMS~\cite{Chatrchyan:2013xna} and ATLAS~\cite{Aad:2014kra}. For $ m_{\tilde{t}} < m_t$ the bounds are taken from \cite{Aad:2014qaa} and for  $m_t < m_{\tilde{t}} < 200$~GeV the bounds are taken from the measurement of spin correlations in $t\bar{t}$ events~\cite{Aad:2014mfk}.
\item $\tilde{g} \to t \bar{t} \nu$.  We study LHC supersymmetry searches for $\tilde{g} \to t \bar{t} \tilde{N}$. The applicable bounds are those for $m_{\tilde{N}} = 0$~GeV. The strongest bounds are derived from the 0-1 leptons and ($\ge 3$) $b$-jets  search from  ATLAS~\cite{Aad:2014lra} and 1 lepton and ($\ge 6$) jets search  from CMS~\cite{Chatrchyan:2013iqa}.
\item $\tilde{g} \to t b b$. We study the  ATLAS RPV search for $\tilde{g} \to t b s $~\cite{Aad:2014pda}, and take the search to be valid for additional $b$-jet multiplicity. 
\end{itemize}

Prompt searches may also apply to LSPs with large enough boosts such that its decay products point back to the PV. Specific displaced tracking algorithms must be used to reconstruct tracks with large transverse impact parameter. Without simulating the full detector and tracking algorithms, it is not known how the prompt search bounds extend for longer lifetimes and we therefore do not consider this possibility.  

\bibliographystyle{ieeetr}

\end{document}